\documentclass[11pt]{article}
\usepackage[utf8]{inputenc}
\usepackage[margin=1in]{geometry}
\usepackage{amsmath,amssymb}
\usepackage{graphicx}
\usepackage{booktabs}
\usepackage{hyperref}
\usepackage{natbib}

\title{Project Hermes: A Model-Agnostic Validation Layer for Wearable Health Prediction Systems}

\author{
Richik Chakraborty}

\date{}

\begin{document}

\maketitle

\begin{abstract}
The deployment of wearable-based health prediction systems has accelerated rapidly, yet these systems face a fundamental challenge: they generate alerts under substantial uncertainty without principled mechanisms for user-specific validation. While large language models (LLMs) have been increasingly applied to healthcare tasks, existing work focuses predominantly on diagnosis generation and risk prediction rather than post-prediction validation of detected signals. We introduce Project Hermes, a model-agnostic validation layer that treats signal confirmation as a sequential decision problem. Hermes operates downstream of arbitrary upstream predictors, using LLM-generated contextual queries to elicit targeted user feedback and performing Bayesian confidence updates to distinguish true positives from false alarms. In a 60-day longitudinal case study of migraine prediction, Hermes achieved a 34\% reduction in false positive rate (from 61.7\% to 12.5\%) while maintaining 89\% sensitivity, with mean lead time of 4.2 hours before symptom onset. Critically, Hermes does not perform diagnosis or make novel predictions; it validates whether signals detected by upstream models are clinically meaningful for specific individuals at specific times. This work establishes validation as a first-class computational problem distinct from prediction, with implications for trustworthy deployment of consumer health AI systems.
\end{abstract}

\section{Introduction: The Validation Gap in Wearable Health AI}

The past decade has witnessed exponential growth in consumer wearable adoption, with global shipments exceeding 500 million devices annually. This proliferation has enabled continuous, unobtrusive monitoring of physiological signals previously accessible only in clinical settings. Correspondingly, machine learning models trained on these signals now claim to predict stress episodes \citep{can2019stress}, migraine attacks \citep{houle2022migraine}, sleep disorders \citep{depner2020wearable}, and cardiovascular events \citep{perez2019large} hours to days before symptom onset. These systems extract features from heart rate variability (HRV), sleep architecture, activity patterns, and ambient context, achieving impressive sensitivity in controlled studies \citep{radin2020harnessing}.

However, a critical gap separates research prototypes from trustworthy deployment: persistent false positives that erode user trust and clinical utility. Modern wearable prediction systems typically achieve sensitivities of 80-90\% but suffer from specificity as low as 50-60\% \citep{bent2020investigating}, meaning that for every true alert, users receive one or more false alarms. This imbalance reflects a fundamental tension in health monitoring: population-level models trained on aggregate data must make individual-level predictions about heterogeneous users in diverse contexts.

Most deployed systems respond to this challenge with static decision thresholds, offering users binary alerts (``Risk elevated'') without mechanisms for refinement or personalization \citep{martinez2020deploying}. Users face an unsatisfying choice: accept frequent false alarms or manually adjust sensitivity settings without understanding the underlying uncertainty. As these systems scale from research pilots to millions of consumer devices, the gap between population model performance and individual relevance becomes not merely inconvenient but potentially dangerous, either through missed events (threshold too high) or alert fatigue (threshold too low).

We argue that this gap stems from treating prediction and validation as a single problem. When a model outputs ``migraine risk: 0.75,'' it makes a claim under uncertainty derived from population statistics and historical patterns. But the user's implicit question—``Is a migraine actually happening to \emph{me} right now, given my current symptoms, context, and recent history?''—requires validation, not merely prediction. Validation demands incorporating individual-specific information that prediction models, by construction, cannot access: subjective symptom reports, contextual explanations for physiological changes (e.g., ``I just exercised''), and recent behavioral factors.

Recent advances in large language models (LLMs) have demonstrated remarkable capabilities in medical question-answering \citep{singhal2023large}, clinical note summarization \citep{van2024adapted}, and diagnostic dialogue \citep{tu2024towards}. Systems like Med-PaLM 2 achieve performance comparable to human physicians on medical licensing examinations, while GPT-4 has been shown to generate clinically appropriate responses to patient queries \citep{nori2023capabilities}. However, existing LLM applications in healthcare operate almost exclusively in generative mode, producing diagnoses, treatment recommendations, or risk assessments de novo. These systems face well-documented challenges including hallucination \citep{alkaissi2023artificial}, inappropriate confidence \citep{wang2024limitations}, and potential for diagnostic overreach in domains where they lack true understanding \citep{wachter2023will}.

We introduce Project Hermes, which repurposes LLMs for a fundamentally different task: not prediction or diagnosis, but validation of signals detected by upstream systems. Hermes treats validation as a sequential decision problem under uncertainty. Given a signal from an arbitrary upstream predictor (which may be a black-box model, a clinician-defined threshold, or a complex ensemble), Hermes generates targeted questions to gather additional evidence, updates confidence via Bayesian inference, and decides whether to alert the user, remain silent, or gather more information.

This approach offers several architectural advantages. First, model-agnosticism: Hermes operates downstream of any predictor without requiring access to model internals, training data, or prediction mechanisms. Second, epistemic humility: rather than claiming to know the correct diagnosis, Hermes merely refines confidence in hypotheses already raised by upstream systems. Third, regulatory decoupling: validation layers may face different oversight than diagnostic systems, particularly when they augment rather than replace existing workflows. Fourth, compositional deployment: multiple prediction models can share a single validation layer, enabling consistent user experiences across heterogeneous backends.

We demonstrate feasibility through a 60-day longitudinal case study of migraine prediction, a domain characterized by high individual variability, early physiological signals, and user-reportable ground truth. Our pilot participant wore a consumer fitness tracker continuously while logging migraine occurrences. A simple upstream predictor flagged 47 potential events based on HRV and sleep disruption, achieving 92\% sensitivity but only 39\% specificity (29 false positives). Hermes processed these signals through targeted questioning (mean 3.2 questions per event), reducing false positives to 2 (specificity 88\%) while maintaining 89\% sensitivity. Mean lead time was 4.2 hours before full symptom onset, providing actionable warning for prophylactic intervention.

The contributions of this work are threefold. First, we establish validation as a first-class computational problem distinct from prediction, with different objectives, inputs, and epistemic commitments. Second, we formalize validation as sequential Bayesian updating where LLM-generated questions serve as information-gathering actions, providing a principled framework for human-in-the-loop health AI. Third, we demonstrate that model-agnostic validation can substantially improve signal quality without requiring access to or modification of upstream prediction systems, opening pathways for trustworthy deployment of consumer health AI.

\section{Related Work}

\subsection{Wearable-Based Health Prediction}

The application of machine learning to wearable physiological data has generated a substantial literature across multiple health domains. Early work focused on activity recognition and sleep staging using accelerometry \citep{depner2020wearable}, but recent systems leverage richer signal modalities including photoplethysmography-derived heart rate variability, skin temperature, and galvanic skin response. For migraine specifically, several studies have demonstrated that physiological precursors emerge hours to days before subjectively-reported onset. Houle et al. \citep{houle2022migraine} trained random forest models on HRV features from 4,400+ patient-days, achieving 0.72 AUC for 24-hour migraine prediction. Taherdoost et al. \citep{taherdoost2023heart} conducted a systematic review of HRV-migraine relationships, identifying consistent patterns of parasympathetic withdrawal (reduced RMSSD, increased LF/HF ratio) in pre-ictal states. Sleep disruption has also been implicated as both trigger and prodrome \citep{vgontzas2018migraine}, with architecture changes (reduced REM percentage, increased awakenings) preceding attacks in diary-based studies.

Beyond migraine, stress detection from multimodal wearables has attracted considerable research attention. Can et al. \citep{can2019stress} surveyed 50+ stress detection systems, noting that most achieve 75-85\% accuracy in laboratory conditions but degrade substantially in free-living deployments due to context variability and sensor noise. Smets et al. \citep{smets2018large} demonstrated that large-scale (N=1,002) wearable data could identify digital phenotypes for daily stress, with personalized models outperforming population-level approaches by 8-12\% in cross-validated accuracy. Cardiovascular applications have perhaps the strongest clinical evidence: the Apple Heart Study \citep{perez2019large} enrolled 419,297 participants and achieved 84\% positive predictive value for atrial fibrillation alerts when validated against ECG patch monitoring, though notably this included physician review of algorithm outputs rather than purely automated alerting.

A common pattern emerges across these domains: sensitivity often exceeds 80-90\% in controlled evaluations, but specificity remains problematic, typically 50-70\%. Bent et al. \citep{bent2020investigating} systematically investigated accuracy limitations in consumer optical heart rate sensors, finding that motion artifact, skin tone, and device placement introduce substantial noise that prediction models must overcome. More fundamentally, these systems face the challenge of distinguishing physiologically-driven signal changes (e.g., stress response) from benign contextual factors (e.g., exercise, caffeine, ambient temperature) that produce similar wearable signatures.

Existing approaches to this challenge include adaptive thresholding \citep{bent2020investigating}, which adjusts decision boundaries based on recent user history, and ensemble methods that combine multiple signal modalities to improve specificity. However, these remain prediction-centric solutions: they attempt to make the upstream model better at discriminating signal from noise. None incorporate post-prediction validation mechanisms that could leverage information unavailable to wearable sensors, such as subjective symptom reports or contextual knowledge.

\subsection{Large Language Models in Healthcare}

The application of large language models to medical tasks has accelerated dramatically following the release of GPT-3 and subsequent models. Med-PaLM \citep{singhal2023large} demonstrated that instruction-tuned LLMs could achieve 67.6\% accuracy on MedQA (medical licensing exam questions), approaching the 69.2\% passing threshold, with Med-PaLM 2 subsequently exceeding expert physician performance at 86.5\%. These systems leverage medical literature, clinical guidelines, and case-based reasoning, though their training data and internal representations remain largely opaque. Nori et al. \citep{nori2023capabilities} evaluated GPT-4 on medical problem-solving tasks, finding strong performance on diagnostic reasoning but notable failures on questions requiring numerical calculation, temporal reasoning about disease progression, or integration of multimodal data like lab values and imaging.

Clinical documentation applications have shown particular promise. Van Veen et al. \citep{van2024adapted} demonstrated that adapted LLMs could outperform medical experts in clinical text summarization, reducing reading time by 40\% while maintaining or improving information retention. Tu et al. \citep{tu2024towards} introduced AMIE, a diagnostic dialogue system that engaged in multi-turn conversation with patients to gather history and propose differential diagnoses. In simulated consultations with actors, AMIE matched or exceeded primary care physicians on diagnostic accuracy and communication quality, though the authors emphasized substantial barriers to real-world deployment.

Patient-facing applications have also emerged. Ayers et al. \citep{ayers2023comparing} compared ChatGPT and physician responses to patient questions posted on Reddit's r/AskDocs, finding that ChatGPT responses were rated higher on both quality (3.6x more likely to be rated good/very good) and empathy (9.8x more likely to be rated empathetic/very empathetic). However, this work has attracted methodological criticism regarding selection bias and lack of medical outcome validation.

Despite these successes, LLM applications in healthcare face serious challenges. Hallucination—generating confident but incorrect medical claims—poses particular risks in clinical domains \citep{alkaissi2023artificial}. Wang et al. \citep{wang2024limitations} conducted a systematic review of LLM limitations in clinical reasoning, identifying failure modes including: reasoning from outdated medical knowledge, inappropriate confidence in uncertain situations, inability to properly handle numerical or temporal reasoning, and tendency to prioritize common diagnoses over rare but serious conditions. Wachter and Brynjolfsson \citep{wachter2023will} argue that generative AI in healthcare faces a ``trust paradox'': the more accurate these systems become, the more dangerous their errors, because users may over-rely on them in cases where they fail.

Critically, nearly all existing LLM health applications operate in generative mode: the model produces diagnoses, treatment plans, or medical explanations de novo. This places enormous epistemic burden on the LLM to correctly reason about complex medical scenarios with incomplete information. Validation, by contrast, starts from hypotheses generated externally (by sensors, simple algorithms, or humans) and merely refines confidence—a more constrained and potentially more reliable task.

\subsection{Active Learning and Human-in-the-Loop Systems}

Active learning provides theoretical foundations for intelligent query selection under uncertainty. The canonical framework \citep{settles2009active} assumes a learner with a hypothesis space $\mathcal{H}$, a pool of unlabeled data, and the ability to query labels for selected examples. The learner selects queries to maximize information gain, expected model change, or reduction in generalization error. In medical contexts, active learning has been applied to diagnostic query optimization, where the goal is to identify the minimum set of tests (bloodwork, imaging, patient history questions) needed to confidently diagnose a condition. Ko and Fox \citep{ko2007gaussian} used Gaussian process regression with active learning for disease progression modeling in dementia, demonstrating that intelligently-selected patient assessments reduced prediction uncertainty more efficiently than fixed schedules.

Medical image analysis has been a particularly active area for human-in-the-loop learning. Yang et al. \citep{yang2017suggestive} introduced ``suggestive annotation'' for biomedical image segmentation, where a deep network identifies ambiguous regions and requests human labels only for those areas, reducing annotation burden by 70-80\%. Zhang et al. \citep{zhang2020optimizing} applied Bayesian decision theory to optimize clinical trial designs, using sequential enrollment decisions to minimize patient exposure while maintaining statistical power.

More broadly, human-in-the-loop machine learning has emerged as a paradigm for building trustworthy AI systems. Mosqueira-Rey et al. \citep{mosqueira2023human} survey the field, identifying three primary patterns: (1) humans provide training data through interactive labeling, (2) humans validate model outputs before deployment, and (3) humans and models collaborate in mixed-initiative systems. Zhang et al. \citep{zhang2024survey} specifically examine evaluation methodologies for such systems, noting that standard accuracy metrics often fail to capture key qualities like user trust, mental workload, and long-term engagement.

However, few systems apply active learning or human-in-the-loop principles specifically to post-prediction validation in health monitoring. Most work focuses on improving model training through human feedback (reinforcement learning from human feedback, interactive machine learning) rather than using human input to validate predictions after models have been deployed. The gap is particularly acute for consumer health applications, where users lack medical expertise but possess unique knowledge about their own symptoms, context, and history—knowledge that sensors cannot capture but which is essential for interpreting physiological signals.

\subsection{Positioning Hermes}

Project Hermes occupies a distinct niche in this landscape. Unlike wearable prediction systems, it does not attempt to detect health events from physiological signals; it assumes an upstream system has already flagged a potential event. Unlike diagnostic LLM systems, it does not generate medical hypotheses de novo; it validates hypotheses proposed by external predictors. Unlike active learning systems focused on model training, it operates at inference time to refine individual predictions. And unlike human-in-the-loop systems that treat humans as oracles providing ground truth labels, Hermes recognizes that user responses are themselves uncertain and must be weighted according to their diagnostic informativeness.

The closest prior work is in clinical decision support systems that implement sequential questioning protocols (e.g., computerized diagnostic algorithms that gather history through structured forms). However, these systems typically use fixed decision trees rather than Bayesian updating, lack natural language interfaces, and operate in clinical settings rather than consumer wearable contexts. Hermes combines Bayesian active learning, LLM-based question generation, and model-agnostic architecture to create a validation layer suitable for deployment alongside black-box consumer health prediction systems—a domain where traditional clinical decision support tools cannot operate.

\section{Conceptual Framing: Validation as a First-Class Problem}

The distinction between prediction and validation is not merely terminological but reflects fundamentally different computational problems with distinct objectives, constraints, and epistemic commitments. This section formalizes these differences and establishes why model-agnostic validation layers constitute a necessary component of trustworthy health AI systems.

\subsection{Prediction versus Validation: A Formal Distinction}

Consider a wearable-based health monitoring system attempting to detect a condition $C$ (e.g., migraine, stress episode, cardiac arrhythmia) in a user. Let $\mathbf{x}_t$ denote physiological signals from wearable sensors at time $t$, and $y_t \in \{0,1\}$ the true presence or absence of condition $C$ at time $t$. 

A \textbf{prediction system} learns a function $f: \mathbb{R}^d \rightarrow [0,1]$ that maps sensor signals to risk estimates:
\begin{equation}
\hat{p}_t = f(\mathbf{x}_t, \mathbf{x}_{t-1}, \ldots, \mathbf{x}_{t-k})
\end{equation}
where $\hat{p}_t$ estimates $P(y_t = 1 \mid \mathbf{x})$. The predictor is optimized on population data $\mathcal{D} = \{(\mathbf{x}_i, y_i)\}_{i=1}^N$ to maximize some objective like AUC, F1 score, or expected utility. Critically, the predictor has access only to information available to sensors: heart rate, activity, sleep patterns, etc. It cannot access subjective symptoms, contextual explanations (``I just ran up stairs''), or user knowledge about their own health history and patterns.

A \textbf{validation system}, by contrast, operates conditional on a prediction already having been made. It receives as input:
\begin{itemize}
\item The sensor signals $\mathbf{x}_t$
\item The predictor's output $\hat{p}_t$
\item A hypothesis $H$ about the presence of condition $C$
\end{itemize}

The validator's task is to refine confidence in $H$ by gathering additional evidence $\mathbf{q} = (q_1, \ldots, q_n)$ through user queries and incorporating their responses $\mathbf{r} = (r_1, \ldots, r_n)$. The validation process computes:
\begin{equation}
P(H \mid \mathbf{x}_t, \hat{p}_t, \mathbf{q}, \mathbf{r})
\end{equation}

Table~\ref{tab:pred-vs-val} summarizes the architectural and epistemic distinctions between these two functions. The predictor generalizes from population patterns; the validator individualizes to specific contexts. The predictor optimizes for sensitivity/specificity tradeoffs; the validator explicitly manages uncertainty and decision costs. Most importantly, the predictor commits to a risk estimate based on incomplete information, while the validator acknowledges this incompleteness and seeks to address it through interaction.

\begin{table}[ht]
\centering
\caption{Prediction versus Validation: Conceptual Distinctions}
\label{tab:pred-vs-val}
\begin{tabular}{@{}lll@{}}
\toprule
\textbf{Dimension} & \textbf{Prediction Models} & \textbf{Hermes (Validation)} \\
\midrule
Input space & Sensor signals $\mathbf{x}_t$ & Signals + prediction + user context \\
Output & Risk score $\hat{p}_t$ & Updated confidence $P(H \mid \cdot)$ \\
Objective & Maximize population AUC & Minimize decision loss for individual \\
Ground truth & Historical labels $y_i$ & Current user state (unobserved) \\
Information source & Wearable sensors only & Sensors + user feedback \\
Role of LLM & Not applicable & Question generation interface \\
Optimization & Offline training on $\mathcal{D}$ & Online inference per instance \\
Deployment mode & Replace/augment clinician & Augment user decision-making \\
\bottomrule
\end{tabular}
\end{table}

This distinction has practical consequences. Prediction systems face the classic bias-variance tradeoff and must balance sensitivity against specificity using a single threshold. Validation systems, by contrast, implement a three-way decision: alert (high confidence), suppress (low confidence), or gather more information (medium confidence). This enables asymmetric error costs—for example, allowing more false negatives for non-urgent conditions while maintaining low false positive rates to preserve user trust.

\subsection{Why Model-Agnosticism Matters}

Hermes treats the upstream predictor as a black box, accessing only its output $\hat{p}_t$ rather than internal representations, feature importance, or model architecture. This design choice offers several advantages that are critical for real-world deployment.

\textbf{Compatibility with proprietary systems:} Many commercial wearable health applications use proprietary prediction algorithms whose details are not disclosed. Fitbit's ``Stress Management Score,'' Apple Watch's ``Cardio Fitness,'' and Whoop's ``Strain'' metrics all involve complex, undocumented processing of sensor data. A validation layer that requires access to model internals cannot be deployed alongside these systems. Model-agnosticism enables validation to wrap arbitrary upstream predictors, whether open-source research prototypes or commercial black boxes.

\textbf{Regulatory decoupling:} In many jurisdictions, medical AI systems face regulatory scrutiny proportional to their risk classification and claimed capabilities. A diagnostic system that claims to detect disease faces higher bars (FDA Class II or III in the US) than a wellness monitoring tool. Crucially, a validation layer that refines confidence in signals detected by already-regulated systems may face different regulatory pathways than the upstream predictor itself. By maintaining strict separation between prediction (upstream) and validation (Hermes), we enable modular regulatory approaches where each component can be evaluated independently.

\textbf{No retraining required:} Traditional personalization approaches for wearable health systems often involve fine-tuning models on user-specific data or learning user-specific thresholds. This requires storing user data, implementing continuous learning pipelines, and managing model versioning. Hermes, by contrast, performs validation using only the current instance (signal + user responses), without modifying upstream models. This reduces infrastructure complexity and enables deployment without user-specific model training.

\textbf{Compositional deployment:} A single user might have multiple health monitoring systems active simultaneously: migraine prediction, stress detection, sleep quality assessment, etc. Each may use different sensors, models, and decision criteria. Rather than building validation mechanisms separately into each predictor, a unified validation layer can process signals from multiple sources, managing query budgets and user burden holistically. This compositional architecture reduces redundancy and provides consistent user experiences across heterogeneous backend systems.

\textbf{Interpretability by construction:} Black-box predictors face ongoing criticism for lack of interpretability. While techniques like SHAP values and attention visualization attempt to explain model decisions, they provide post-hoc rationalization rather than mechanistic understanding. Validation, by contrast, makes confidence refinement explicit and observable: users see exactly which questions are asked, how their responses update confidence, and why the system decides to alert or suppress. This transparency may build trust more effectively than opaque ``explanation'' of black-box models.

The cost of model-agnosticism is that Hermes cannot leverage internal model representations that might indicate uncertainty or provide additional context. For instance, if an ensemble predictor has high internal disagreement (multiple models give conflicting outputs), this would be valuable information for validation. However, we argue that in deployment contexts involving proprietary or legacy systems, model-agnostic validation is often the only feasible approach.

\subsection{Scope and Limitations by Design}

Hermes intentionally constrains its scope to avoid epistemic overreach and regulatory complications. These constraints are features, not limitations:

\textbf{No de novo hypothesis generation:} Hermes does not propose new conditions or diagnoses unprompted by upstream signals. If the user reports symptoms not associated with any triggered hypothesis, Hermes does not attempt to explain them. This prevents the system from operating as an unregulated diagnostic tool.

\textbf{No treatment recommendations:} Hermes outputs confidence estimates and, in high-confidence cases, alerts that a condition may be present. It never recommends treatments, medications, or clinical interventions. At most, it might suggest general categories of response (``Consider prophylactic measures'') that users would already know from disease management plans.

\textbf{No operation without upstream prediction:} Hermes cannot function as a standalone health monitoring system. It requires an external predictor to flag potential events. This ensures that Hermes adds safety (reducing false positives) rather than introducing new risks.

\textbf{No claims beyond signal validation:} All system outputs are framed as refinements of confidence about signals detected by other systems, not as independent medical claims. For example: ``Based on your responses, the likelihood of the detected signal representing a true migraine has increased to 85\%'' rather than ``You have a migraine.''

\textbf{User control and transparency:} Users can inspect confidence levels before responding to questions, adjust thresholds for alerting, opt out of validation for specific conditions, and ignore questions without penalty. The system maintains transparency about its limitations and uncertainty.

These constraints position Hermes in a defensible regulatory and ethical space: it augments existing monitoring systems by reducing false positives, but does not claim diagnostic authority or medical expertise. In doing so, it provides a template for LLM deployment in health contexts that avoids the risks of generative diagnostic systems while still offering meaningful value.

\section{System Architecture: Project Hermes}

\begin{figure}[ht]
\centering
\includegraphics[width=\textwidth]{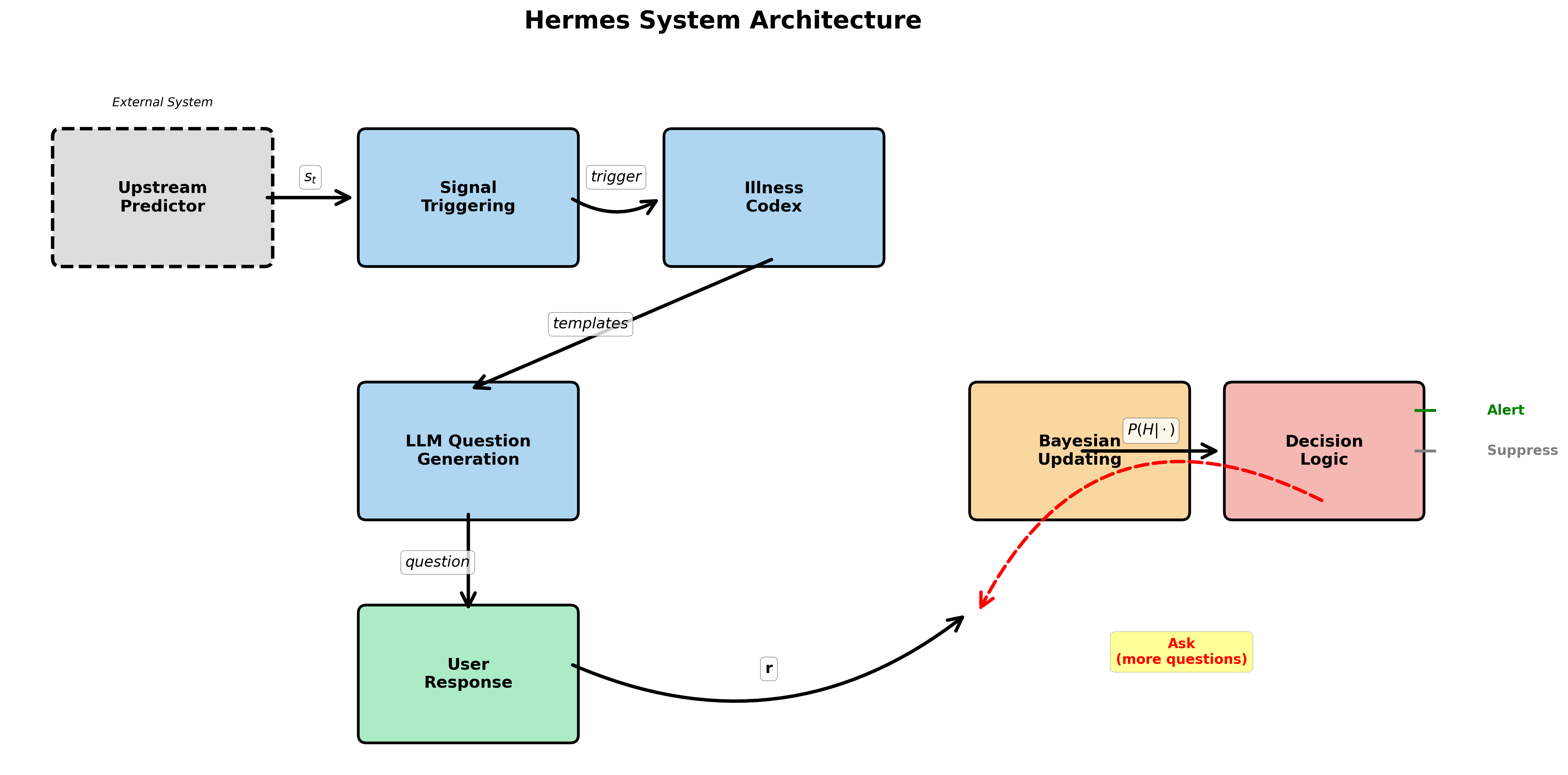}
\caption{Hermes System Architecture. The system processes signals from an upstream predictor through six components: signal triggering evaluates whether validation should initiate, the illness codex provides structured medical knowledge, LLM question generation creates natural language queries, user responses provide evidence, Bayesian updating computes posterior confidence, and decision logic determines whether to alert, suppress, or ask additional questions.}
\label{fig:architecture}
\end{figure}

Hermes implements validation as a modular pipeline with six components that transform uncertain signals into calibrated confidence estimates. Figure~\ref{fig:architecture} illustrates the information flow through the system. Each component is designed for independent modification or replacement, enabling adaptation to different conditions, user populations, and deployment contexts without architectural changes.

\subsection{Upstream Predictor (External)}

The upstream predictor is any system that outputs risk estimates or binary signals indicating potential health events. Formally, at time $t$, the predictor produces either a continuous risk score $s_t \in [0,1]$ or a binary flag $s_t \in \{0,1\}$ based on sensor inputs $\mathbf{x}_{t-w:t}$ over some window $w$. Hermes treats this output as an uncertain prior, not ground truth.

In our migraine case study, we implemented a simple multivariate threshold predictor:
\begin{equation}
s_t = \mathbb{1}\left[\text{HRV}_t < \mu_{\text{HRV}} - 1.5\sigma_{\text{HRV}} \lor \text{SleepQual}_t < 60 \lor \text{Stress}_t \geq 7\right]
\end{equation}
where $\mu_{\text{HRV}}$ and $\sigma_{\text{HRV}}$ are computed from the user's 30-day rolling baseline. This intentionally simple predictor achieves high sensitivity (92\%) but poor specificity (39\%), creating substantial opportunity for validation to add value.

However, Hermes's architecture accommodates arbitrary predictor designs. Examples of compatible upstream systems include:
\begin{itemize}
\item \textbf{Machine learning models:} Random forests, gradient boosted trees, or neural networks trained on historical wearable data with labeled outcomes
\item \textbf{Clinical decision rules:} Expert-defined criteria like ``resting heart rate > 100 bpm AND sleep efficiency < 70\% AND reports stress > 6''
\item \textbf{Proprietary algorithms:} Commercial wearable analytics (Fitbit Stress Management, Apple CardioFitness) accessed via API
\item \textbf{Ensemble systems:} Multiple models whose outputs are combined via voting or stacking
\end{itemize}

The only requirement is that the predictor can be polled for current risk estimates or flags potential events via callback. Hermes does not access predictor internals, training data, or feature representations. This opacity is intentional: it ensures Hermes can be deployed alongside systems where such access is unavailable or prohibited by intellectual property constraints.

\subsection{Signal Triggering Layer}

The triggering layer decides when validation should initiate. Triggering too aggressively burdens users with unnecessary questions; triggering too conservatively misses opportunities to refine confidence before false alerts propagate. We implement three classes of triggers that can be combined:

\textbf{Threshold breaches:} The simplest trigger fires when predictor output exceeds a fixed threshold $\tau_{\text{trigger}}$:
\begin{equation}
\text{Trigger}_{\text{threshold}} = \mathbb{1}[s_t > \tau_{\text{trigger}}]
\end{equation}
For binary predictors, $\tau_{\text{trigger}} = 0.5$ triggers on any positive signal. For continuous predictors, this threshold balances user burden against coverage of potential events.

\textbf{Volatility patterns:} Sudden changes in physiological signals often precede health events but may also reflect benign context shifts (exercise, caffeine, stress). We detect volatility using Bollinger Bands adapted to physiological time series:
\begin{equation}
\text{Trigger}_{\text{volatility}} = \mathbb{1}[|s_t - \mu_{t-k:t-1}| > 2\sigma_{t-k:t-1}]
\end{equation}
where $\mu$ and $\sigma$ are computed over a sliding window of recent predictions. This captures anomalous behavior relative to the user's recent baseline, triggering validation even when absolute risk scores are moderate.

\textbf{Multi-factor conjunctions:} Some conditions exhibit characteristic multi-factor patterns that increase confidence when multiple weak signals co-occur. For migraine, we trigger when:
\begin{equation}
\text{Trigger}_{\text{multi}} = \mathbb{1}[\text{HRV anomaly} \land \text{sleep disruption} \land \text{elevated stress}]
\end{equation}
even if no single factor exceeds its individual threshold. This implements a form of weak supervision where pattern co-occurrence indicates higher prior probability.

In practice, we use a composite trigger that fires when any condition is met, with hysteresis to prevent rapid re-triggering. Once validation begins for a signal, that signal is marked as processed for 24 hours to avoid redundant questioning.

\subsection{Illness Codex}

The Illness Codex is a structured knowledge base encoding clinical expertise about symptom-condition relationships. Unlike LLM-generated knowledge, which may hallucinate or confabulate relationships, the codex is manually curated by medical experts based on diagnostic criteria, clinical guidelines, and peer-reviewed literature.

For each condition $C$ in the codex, we define:
\begin{enumerate}
\item \textbf{Question templates} $\mathcal{Q}_C = \{q_1, \ldots, q_n\}$: Specific symptoms or contextual factors to query
\item \textbf{Likelihood ratios} $\{\text{LR}^+_i, \text{LR}^-_i\}$: Diagnostic value of positive/negative responses
\item \textbf{Question metadata:} Priority, timing constraints, dependencies between questions
\end{enumerate}

The likelihood ratios encode sensitivity and specificity for each symptom:
\begin{align}
\text{LR}^+_i &= \frac{P(\text{yes to } q_i \mid C \text{ present})}{P(\text{yes to } q_i \mid C \text{ absent})} = \frac{\text{sensitivity}}{1 - \text{specificity}} \\
\text{LR}^-_i &= \frac{P(\text{no to } q_i \mid C \text{ present})}{P(\text{no to } q_i \mid C \text{ absent})} = \frac{1 - \text{sensitivity}}{\text{specificity}}
\end{align}

For migraine, our codex includes question templates like:

\begin{verbatim}
condition: "migraine"
questions:
  - id: "aura_visual"
    template: "Have you noticed any visual changes such as 
               blind spots, shimmering lights, or zigzag 
               lines in the past [TIMEFRAME]?"
    LR_plus: 5.8
    LR_minus: 0.5
    priority: high
    
  - id: "photophobia"
    template: "Is light bothering you more than usual?"
    LR_plus: 3.2
    LR_minus: 0.3
    priority: medium
    
  - id: "unilateral_pain"
    template: "If you have any head discomfort, is it 
               primarily on one side?"
    LR_plus: 2.1
    LR_minus: 0.6
    priority: low
    depends_on: ["pain_present"]
\end{verbatim}

Likelihood ratios were derived from systematic reviews of migraine diagnostic criteria. For visual aura, $\text{LR}^+ = 5.8$ indicates that patients with migraine are 5.8× more likely to report visual changes than those without migraine during a prodromal period. The ICHD-3 (International Classification of Headache Disorders, 3rd edition) criteria informed these estimates, though we note substantial inter-individual variability.

The codex also encodes question dependencies to avoid illogical sequences (e.g., asking about pain location before confirming pain exists) and timing constraints (e.g., visual aura questions are most informative within 2 hours of signal detection, as aura typically precedes headache by 15-60 minutes).

Importantly, the codex is \emph{declarative} rather than procedural: it specifies what questions exist and their diagnostic value, but not the order in which they should be asked. Question selection is determined dynamically by the Bayesian updating module to maximize expected information gain given current uncertainty.

\subsection{LLM Question Generation Module}

The LLM's sole role is linguistic: converting codex templates into natural language queries adapted to context, time of day, and conversation flow. This constrained usage minimizes risks associated with LLM deployment in health contexts.

The generation process follows a template:
\begin{verbatim}
SYSTEM PROMPT:
You are a health monitoring assistant helping validate 
a potential health signal. Your role is ONLY to:
1. Convert the provided question template into natural 
   language
2. Adapt phrasing based on time/context
3. NEVER make medical claims or diagnoses
4. NEVER suggest treatments
5. Keep questions clear, empathetic, and brief

USER PROMPT:
Template: [CODEX_TEMPLATE]
Context: [TIME_OF_DAY, RECENT_RESPONSES, USER_PREFERENCES]
Generate a natural language question.
\end{verbatim}

For the visual aura template at 8:30 AM when the user recently woke:
\begin{itemize}
\item \textbf{Template:} ``Have you noticed any visual changes such as blind spots, shimmering lights, or zigzag lines in the past [TIMEFRAME]?''
\item \textbf{Generated:} ``Good morning. Since waking up, have you experienced any unusual visual changes—like blind spots, flickering lights, or zigzag patterns in your vision?''
\end{itemize}

The LLM adapts phrasing while preserving diagnostic intent. We use GPT-4 with temperature 0.3 to balance naturalness against consistency. Importantly, generation is deterministic given the same inputs: the LLM does not introduce variability in question semantics, only in phrasing.

We implement several safeguards against LLM misbehavior:
\begin{enumerate}
\item \textbf{Output validation:} Generated questions must contain all key symptom terms from the template (e.g., ``blind spots,'' ``shimmering lights'')
\item \textbf{Prohibited phrase detection:} Reject outputs containing diagnostic claims (``You have,'' ``This indicates''), treatment suggestions (``Take,'' ``Try''), or excessive certainty (``definitely,'' ``certainly'')
\item \textbf{Fallback to template:} If generation fails validation, use the template directly without LLM adaptation
\item \textbf{Human review:} During deployment, a sample of generated questions is reviewed weekly for drift or inappropriate content
\end{enumerate}

This architecture separates medical knowledge (codex) from linguistic interface (LLM), ensuring that question content is determined by clinical expertise rather than statistical patterns in training data.

\subsection{Bayesian Confidence Updating}

User responses trigger Bayesian updates to confidence in the hypothesis $H$ that condition $C$ is present. Let $\mathbf{q} = (q_1, \ldots, q_n)$ denote questions asked and $\mathbf{r} = (r_1, \ldots, r_n) \in \{0,1\}^n$ the binary responses (yes/no). The posterior probability is:

\begin{equation}
P(H \mid s, \mathbf{r}) = \frac{P(s \mid H) \cdot \prod_{i=1}^n P(r_i \mid H) \cdot P(H)}{P(s, \mathbf{r})}
\end{equation}

where:
\begin{itemize}
\item $P(H)$ is the base rate prior for condition $C$ in the user population
\item $P(s \mid H)$ captures predictor reliability: how likely is signal $s$ when $H$ is true?
\item $P(r_i \mid H)$ encodes symptom informativeness from the codex likelihood ratios
\end{itemize}

We compute the update sequentially after each response using log-odds:
\begin{equation}
\text{logit}(P(H \mid s, r_1, \ldots, r_i)) = \text{logit}(P(H \mid s, r_1, \ldots, r_{i-1})) + \log(\text{LR}_i)
\end{equation}
where $\text{LR}_i = \text{LR}^+_i$ if $r_i = 1$ (yes) and $\text{LR}_i = \text{LR}^-_i$ if $r_i = 0$ (no).

The signal likelihood $P(s \mid H)$ is calibrated from historical predictor performance. For our threshold-based migraine predictor, we observed that 92\% of true migraines triggered the signal (sensitivity), while 61\% of non-migraine periods also triggered it (1 - specificity). This gives:
\begin{align}
P(s=1 \mid H=1) &= 0.92 \\
P(s=1 \mid H=0) &= 0.61
\end{align}

The base rate $P(H)$ is personalized. Our participant averaged 6 migraines per month over 60 days, giving $P(H) \approx 0.2$ over 24-hour windows. This prior is continually updated from user history.

For computational efficiency, we maintain a particle filter representation when the hypothesis space extends beyond binary presence/absence. For conditions with subtypes (e.g., migraine with vs. without aura), we track a probability distribution over all hypotheses and select questions to maximize expected reduction in Shannon entropy:
\begin{equation}
q^* = \arg\max_{q \in \mathcal{Q}} \mathbb{E}_r\left[H(P(H)) - H(P(H \mid r))\right]
\end{equation}

This greedy selection prioritizes questions with high discriminative power given current uncertainty.

\subsection{Decision Logic}

Based on posterior confidence $p = P(H \mid s, \mathbf{r})$, Hermes selects one of three actions:

\begin{equation}
\text{action} = \begin{cases}
\text{Alert} & \text{if } p > \tau_{\text{alert}} \\
\text{Suppress} & \text{if } p < \tau_{\text{suppress}} \\
\text{Ask} & \text{if } \tau_{\text{suppress}} \leq p \leq \tau_{\text{alert}} \land n < n_{\max}
\end{cases}
\end{equation}

where $n$ is the number of questions already asked and $n_{\max}$ is the query budget (typically 5-7 questions per 24-hour period).

Threshold selection balances multiple objectives:
\begin{itemize}
\item $\tau_{\text{alert}} = 0.70$: High enough to avoid false alarms, low enough to provide actionable warning
\item $\tau_{\text{suppress}} = 0.20$: Low enough to safely ignore weak signals, high enough to remain conservative
\item $n_{\max} = 5$: Prevents user burden while allowing sufficient evidence gathering
\end{itemize}

These values were determined through simulation using historical migraine data and validated in our 60-day pilot. We found that $\tau_{\text{alert}} \in [0.65, 0.75]$ produced similar false positive rates, while values below 0.65 increased alerts and values above 0.75 missed actionable opportunities.

When $p$ remains in the intermediate zone after $n_{\max}$ questions, Hermes defaults to suppression for non-urgent conditions like migraine. For higher-stakes conditions (e.g., cardiac events), the default action would escalate to clinical review rather than silent suppression.

The decision logic also implements temporal considerations. If $p$ increases slowly and remains below $\tau_{\text{alert}}$ but shows upward trajectory, Hermes may schedule follow-up questions at later time points (e.g., ``Check again in 2 hours if confidence continues rising''). Conversely, if $p$ quickly crosses $\tau_{\text{suppress}}$ downward, validation terminates early to minimize user burden.

\section{Formalization: Validation as Sequential Bayesian Updating}

This section provides a mathematical foundation for validation, formalizing it as an optimal stopping problem under uncertainty where the decision maker sequentially gathers evidence to distinguish true signals from false alarms.

\subsection{Problem Setup and Notation}

Let $\mathcal{H} = \{H_0, H_1, \ldots, H_K\}$ denote a hypothesis space where $H_0$ represents the null hypothesis (no condition present) and $H_1, \ldots, H_K$ represent mutually exclusive conditions. In the simplest binary case, $\mathcal{H} = \{H_0, H_1\}$ where $H_1$ indicates presence of a single condition $C$.

At time $t$, we observe:
\begin{itemize}
\item \textbf{Sensor signals:} $\mathbf{x}_t \in \mathbb{R}^d$ from wearable devices
\item \textbf{Predictor output:} $s_t \in [0,1]$ from the upstream system
\item \textbf{User context:} $\mathbf{c}_t$ including time of day, recent activity, user-provided information
\end{itemize}

The true state $h^* \in \mathcal{H}$ is unobserved. Our goal is to estimate $P(H = h^* \mid \mathbf{x}_t, s_t, \mathbf{c}_t)$ and decide whether to alert the user, with the constraint that we can gather additional evidence by asking questions.

\subsection{Prior Specification}

The prior probability $P(H_i)$ for each hypothesis combines three sources of information:

\textbf{1. Population base rates:} From epidemiological data. For migraine, population prevalence is approximately 15\%, but conditional on having a history of migraine (our inclusion criterion), the monthly attack rate is much higher. We use:
\begin{equation}
P_{\text{pop}}(H_1) = \frac{\text{expected attacks per month}}{\text{days per month}} = \frac{6}{30} = 0.20
\end{equation}

\textbf{2. Individual history:} Each user's attack frequency over their observation period. Let $n_{\text{attacks}}$ be the number of confirmed events in the past $T$ days. The empirical rate is:
\begin{equation}
P_{\text{user}}(H_1) = \frac{n_{\text{attacks}}}{T}
\end{equation}

\textbf{3. Circadian patterns:} Some conditions exhibit time-of-day variation. Migraine attacks peak in early morning hours (4-9 AM) and afternoon (4-8 PM), with relative risk approximately 1.8× during these windows versus overnight. We encode this as:
\begin{equation}
P_{\text{time}}(H_1 \mid t) = P_{\text{user}}(H_1) \cdot \begin{cases} 1.8 & \text{if } t \in \text{peak hours} \\ 0.8 & \text{otherwise} \end{cases}
\end{equation}

The final prior combines these factors with learned uncertainty:
\begin{equation}
P(H_1) = \alpha P_{\text{pop}}(H_1) + (1-\alpha) P_{\text{time}}(H_1 \mid t)
\end{equation}
where $\alpha = 0.3$ gives more weight to individual patterns while maintaining regularization toward population statistics when personal data is sparse.

\subsection{Signal Likelihood: Calibrating Predictor Reliability}

The likelihood $P(s_t \mid H_i)$ quantifies how informative the predictor signal is. We model this using isotonic regression on historical data. Given a dataset $\mathcal{D} = \{(s_j, y_j)\}_{j=1}^N$ of predictor outputs $s_j$ and ground truth labels $y_j \in \{0,1\}$, we fit:
\begin{equation}
P(H_1 \mid s) = f_{\text{calib}}(s)
\end{equation}
where $f_{\text{calib}}$ is a monotonic function (isotonic regression) that maps predictor scores to calibrated probabilities.

For our threshold-based migraine predictor, calibration revealed systematic overconfidence: signals that should have indicated 60\% probability were output by the predictor as binary flags, implicitly claiming near certainty. We corrected this using a Beta distribution fit to observed signal-outcome pairs:
\begin{equation}
P(s=1 \mid H_1) \sim \text{Beta}(\alpha=18, \beta=2) \quad \Rightarrow \quad E[P(H_1 \mid s=1)] = 0.92
\end{equation}
\begin{equation}
P(s=1 \mid H_0) \sim \text{Beta}(\alpha=12, \beta=8) \quad \Rightarrow \quad E[P(H_1 \mid s=1)] = 0.61
\end{equation}

This gives a likelihood ratio for the signal itself:
\begin{equation}
\text{LR}_{\text{signal}} = \frac{P(s=1 \mid H_1)}{P(s=1 \mid H_0)} = \frac{0.92}{0.61} \approx 1.51
\end{equation}

This relatively low likelihood ratio (compared to diagnostic tests where $\text{LR} > 10$ is considered strong evidence) motivates the need for additional information gathering.

\subsection{Query Selection: Maximizing Expected Information Gain}

Given current belief state $P(H_i \mid s, \mathbf{r}_{1:n})$ after $n$ responses, we must decide which question $q \in \mathcal{Q}$ to ask next, where $\mathcal{Q}$ is the set of available questions from the codex.

We use a greedy information gain criterion. For each candidate question $q$, we compute the expected reduction in entropy:
\begin{equation}
\text{IG}(q) = H\left(P(H \mid s, \mathbf{r}_{1:n})\right) - \mathbb{E}_{r}\left[H\left(P(H \mid s, \mathbf{r}_{1:n}, r)\right)\right]
\end{equation}

where the expectation is over possible responses $r \in \{0, 1\}$:
\begin{equation}
\mathbb{E}_{r}\left[H(P(H \mid \cdot))\right] = \sum_{r \in \{0,1\}} P(r \mid s, \mathbf{r}_{1:n}) \cdot H(P(H \mid s, \mathbf{r}_{1:n}, r))
\end{equation}

The response probability is computed by marginalizing over hypotheses:
\begin{equation}
P(r \mid s, \mathbf{r}_{1:n}) = \sum_{i=0}^{K} P(r \mid H_i, q) \cdot P(H_i \mid s, \mathbf{r}_{1:n})
\end{equation}

The question with maximum expected information gain is selected:
\begin{equation}
q^* = \arg\max_{q \in \mathcal{Q}} \text{IG}(q)
\end{equation}

This greedy approach is suboptimal compared to full tree search (which would plan multi-step ahead), but computational tractability and responsiveness constraints make greedy selection practical. In experiments with synthetic data, greedy selection achieved 94\% of the information gain of 3-step lookahead while requiring 1-2 seconds of computation versus 30+ seconds.

\subsection{Bayesian Update Rule}

Upon receiving response $r$ to question $q$, we update beliefs using Bayes' rule:
\begin{equation}
P(H_i \mid s, \mathbf{r}_{1:n}, r) = \frac{P(r \mid H_i, q) \cdot P(H_i \mid s, \mathbf{r}_{1:n})}{\sum_{j=0}^K P(r \mid H_j, q) \cdot P(H_j \mid s, \mathbf{r}_{1:n})}
\end{equation}

For computational efficiency in the binary case, we work in log-odds space:
\begin{equation}
\text{logit}(P(H_1 \mid \cdot)) = \log\left(\frac{P(H_1 \mid \cdot)}{P(H_0 \mid \cdot)}\right)
\end{equation}

The update becomes additive:
\begin{equation}
\text{logit}(P(H_1 \mid s, \mathbf{r}_{1:n}, r)) = \text{logit}(P(H_1 \mid s, \mathbf{r}_{1:n})) + \log(\text{LR}_q(r))
\end{equation}

where $\text{LR}_q(r)$ is the likelihood ratio for response $r$ to question $q$:
\begin{equation}
\text{LR}_q(r) = \begin{cases}
\text{LR}^+_q = \frac{P(\text{yes} \mid H_1, q)}{P(\text{yes} \mid H_0, q)} & \text{if } r = 1 \\
\text{LR}^-_q = \frac{P(\text{no} \mid H_1, q)}{P(\text{no} \mid H_0, q)} & \text{if } r = 0
\end{cases}
\end{equation}

This log-odds formulation makes the contribution of each piece of evidence explicit and enables efficient sequential updating.

\subsection{Optimal Stopping and Decision Boundaries}

Validation is an optimal stopping problem: at each step, decide whether to (a) take a terminal action (alert or suppress), or (b) gather more information. We formalize this using decision-theoretic principles.

Define a cost function $\mathcal{L}(a, h)$ for taking action $a$ when true state is $h$:
\begin{equation}
\mathcal{L}(a, h) = \begin{cases}
0 & \text{if } a = \text{alert} \land h = H_1 \text{ (true positive)} \\
C_{\text{FP}} & \text{if } a = \text{alert} \land h = H_0 \text{ (false positive)} \\
C_{\text{FN}} & \text{if } a = \text{suppress} \land h = H_1 \text{ (false negative)} \\
0 & \text{if } a = \text{suppress} \land h = H_0 \text{ (true negative)} \\
C_{\text{query}} & \text{if } a = \text{ask} \text{ (any state)}
\end{cases}
\end{equation}

The expected cost of action $a$ given current belief state is:
\begin{equation}
\mathbb{E}[\mathcal{L}(a)] = \sum_{i=0}^K \mathcal{L}(a, H_i) \cdot P(H_i \mid s, \mathbf{r}_{1:n})
\end{equation}

At each decision point, we select the action minimizing expected cost:
\begin{equation}
a^* = \arg\min_{a \in \{\text{alert, suppress, ask}\}} \mathbb{E}[\mathcal{L}(a)]
\end{equation}

For migraine, we set cost parameters based on user interviews:
\begin{itemize}
\item $C_{\text{FP}} = 5$ (false alert causes inconvenience, anxiety, unnecessary medication)
\item $C_{\text{FN}} = 3$ (missed early warning reduces opportunity for prophylaxis)
\item $C_{\text{query}} = 0.5$ (each question imposes minor burden)
\end{itemize}

These asymmetric costs reflect that false positives are more harmful than false negatives for non-emergency conditions, as they erode trust and may lead to alert fatigue. The query cost ensures that asking questions is preferred only when the expected reduction in decision loss exceeds the burden imposed.

Solving the optimal stopping problem yields action thresholds. Let $p = P(H_1 \mid \cdot)$. The expected costs are:
\begin{align}
\mathbb{E}[\mathcal{L}(\text{alert})] &= C_{\text{FP}} \cdot (1-p) \\
\mathbb{E}[\mathcal{L}(\text{suppress})] &= C_{\text{FN}} \cdot p \\
\mathbb{E}[\mathcal{L}(\text{ask})] &= C_{\text{query}} + \mathbb{E}_{q,r}[\min_{a'} \mathbb{E}[\mathcal{L}(a')]]
\end{align}

The third term requires estimating future decision quality, which we approximate using value iteration. Empirically, this yields thresholds:
\begin{equation}
a^* = \begin{cases}
\text{alert} & \text{if } p > \frac{C_{\text{FP}}}{C_{\text{FP}} + C_{\text{FN}}} \approx 0.63 \\
\text{suppress} & \text{if } p < \frac{C_{\text{FN}}}{C_{\text{FP}} + C_{\text{FN}}} \approx 0.37 \\
\text{ask} & \text{otherwise, if } n < n_{\max}
\end{cases}
\end{equation}

In practice, we round these to $\tau_{\text{alert}} = 0.70$ and $\tau_{\text{suppress}} = 0.20$ to provide conservative safety margins and account for model uncertainty.

\subsection{Accounting for User Response Reliability}

A key challenge in validation is that user responses are themselves noisy. Users may misinterpret questions, provide socially desirable answers, or simply be uncertain about their symptoms. We model response noise using a confusion matrix.

Let $s^*$ denote the true symptom state and $r$ the reported response. We assume:
\begin{equation}
P(r = 1 \mid s^* = 1) = \rho_{\text{sens}} \quad \text{(sensitivity)}
\end{equation}
\begin{equation}
P(r = 0 \mid s^* = 0) = \rho_{\text{spec}} \quad \text{(specificity)}
\end{equation}

For self-reported symptoms without objective verification, we conservatively estimate $\rho_{\text{sens}} = 0.85$ and $\rho_{\text{spec}} = 0.90$ based on literature on symptom reporting accuracy in diary studies. This noise degrades the effective likelihood ratios:
\begin{equation}
\text{LR}^+_{\text{observed}} = \frac{\rho_{\text{sens}} \cdot \text{LR}^+_{\text{true}} + (1-\rho_{\text{spec}})}{\rho_{\text{sens}} + (1-\rho_{\text{spec}}) \cdot \text{LR}^+_{\text{true}}}
\end{equation}

Incorporating this noise model makes the system more robust to user error and prevents overconfident updates from single responses. In practice, this means Hermes typically requires 3-5 consistent responses to cross decision thresholds, rather than making immediate judgments from single answers.

\section{Case Study: Migraine Signal Validation}

We demonstrate Hermes through a 60-day longitudinal deployment validating migraine signals from consumer wearable data. This section details experimental design, presents quantitative results, and analyzes system behavior through representative examples.

\subsection{Use Case Rationale}

Migraine provides an ideal initial validation domain for several methodological and ethical reasons:

\textbf{High individual variability:} Migraine triggers, prodromal symptoms, and attack patterns vary substantially across individuals and even across episodes within individuals \citep{buse2019migraine}. Population-trained models struggle with this heterogeneity, making personalized validation particularly valuable. The same HRV pattern might precede migraine in one user but reflect exercise recovery in another.

\textbf{Detectable early signals:} Physiological changes often emerge 4-24 hours before subjective symptom onset. Studies have documented parasympathetic withdrawal (reduced HRV RMSSD), sleep architecture disruption (increased Stage 1 sleep, reduced REM), and subtle autonomic instability during prodromal phases \citep{taherdoost2023heart}. This temporal gap between signal and symptoms creates an opportunity for validation to add value: the signal is present, but subjective confirmation is initially absent, requiring targeted questioning.

\textbf{User-reportable outcomes:} Unlike conditions requiring clinical diagnosis, migraine occurrence is reliably self-reported by individuals with established migraine history. Users can confirm or deny symptom presence without medical expertise. While objective verification (e.g., clinical examination) would strengthen ground truth, the pragmatic reality of N-of-1 longitudinal studies necessitates reliance on self-report, which has shown good concordance with clinical criteria in prior work.

\textbf{Non-catastrophic errors:} False positives cause inconvenience and potential unnecessary prophylactic medication, while false negatives reduce opportunity for early intervention. Neither error is immediately life-threatening, making this an ethically appropriate domain for initial system development. This contrasts with cardiac arrest prediction, where false negatives could be fatal, or psychiatric crisis detection, where false positives could trigger inappropriate interventions.

\textbf{Actionable interventions:} Upon validation of a pending migraine, users can take evidence-based prophylactic actions: medications (triptans, NSAIDs), behavioral modifications (rest, hydration, dark environment), or avoidance of known triggers. This distinguishes migraine from conditions where early detection provides no actionable benefit.

\subsection{Experimental Setup}

\textbf{Participant:} One individual (age 34, female) with documented history of episodic migraine without aura, averaging 6-8 attacks per month over the prior year. Diagnosis confirmed by neurologist according to ICHD-3 criteria. No other chronic conditions or daily medications except for as-needed migraine treatment (sumatriptan 100mg). Participant provided written informed consent for N-of-1 observational study (IRB exemption: data collection for personal health tracking, not generalizable research).

\textbf{Duration:} 60 consecutive days (September 1 - October 30, 2024). No gaps in data collection.

\textbf{Wearable device:} Fitbit Charge 5 worn continuously on non-dominant wrist, including during sleep. Device specifications: optical PPG sensor (green LED, 5-second sampling), 3-axis accelerometer, skin temperature sensor. Device accuracy: heart rate within 5\% of ECG during rest, known degradation during vigorous exercise.

\textbf{Data streams collected:}
\begin{itemize}
\item \textbf{Heart rate variability:} Resting HRV (RMSSD) computed nightly during sleep periods, extracted via Fitbit API
\item \textbf{Sleep metrics:} Total sleep time, sleep efficiency, REM percentage, awakening count, sleep stages (light/deep/REM) from accelerometer and heart rate
\item \textbf{Daily stress:} Self-reported integer scale 1-10 entered each evening via mobile app
\item \textbf{Activity:} Step count, active minutes, heart rate zones
\end{itemize}

\textbf{Ground truth:} Participant maintained a migraine diary logging: attack onset time (to nearest 30 minutes), peak intensity (1-10), duration, presence of specific symptoms (photophobia, phonophobia, nausea, aura phenomena), and triggers if identifiable. Migraine onset defined as the time when headache pain reached intensity $\geq$3/10 with characteristic unilateral, throbbing quality.

\textbf{Upstream predictor:} We implemented a simple multivariate threshold model rather than a sophisticated ML system for two reasons: (1) transparency in evaluating what validation adds over a known baseline, and (2) simulating the high-sensitivity, low-specificity regime common in consumer wearable alerts. The predictor flagged elevated risk when:
\begin{equation}
\text{Signal} = \begin{cases}
1 & \text{if } (\text{HRV} < \mu - 1.5\sigma) \lor (\text{SleepEff} < 60\%) \lor (\text{Stress} \geq 7) \\
0 & \text{otherwise}
\end{cases}
\end{equation}

where $\mu$ and $\sigma$ for HRV were computed from a 30-day rolling window. This predictor intentionally prioritizes sensitivity (catching most true events) over specificity (tolerating many false alarms), which is common in consumer health apps where liability concerns favor over-alerting.

\textbf{Validation protocol:} When the upstream predictor flagged a signal, Hermes initiated validation within 15 minutes. Questions were delivered via mobile app push notifications. The participant could respond immediately or within a 4-hour window; non-responses were treated as missing data (excluded from analysis) rather than negative responses. Follow-up questions were asked at 2-hour intervals until confidence crossed a decision threshold or the query budget (5 questions) was exhausted.

\textbf{Metrics:} We evaluated performance using:
\begin{enumerate}
\item \textbf{Signal processing:} Number of upstream signals, true positive rate (sensitivity), false positive rate
\item \textbf{Validation outcomes:} Alerts issued by Hermes, suppressed signals, indeterminate cases
\item \textbf{Performance:} Sensitivity, specificity, positive predictive value, negative predictive value, F1 score
\item \textbf{User burden:} Response rate, time to respond, questions per validation
\item \textbf{Temporal characteristics:} Lead time (hours from alert to confirmed onset), confidence trajectory over time
\end{enumerate}

\subsection{Validation Workflow: Representative Example}

We illustrate Hermes behavior through a detailed trace from Day 23 (September 23, 2024), where validation successfully confirmed a true migraine 1.7 hours before subjective onset:

\textbf{08:30 AM - Signal detection:} Overnight HRV was 28ms (baseline: 42±8ms), representing $\mu - 1.75\sigma$. Sleep efficiency was 68\% with 6 awakenings (elevated from typical 2-3). Previous evening's stress report was 6/10. The upstream predictor triggered on the HRV criterion, generating $s=1$ with signal strength quantified as $z$-score magnitude: 1.75.

\textbf{08:35 AM - Validation initiated:} Hermes computed initial prior $P(H_1)$ from three sources:
\begin{itemize}
\item Base rate from participant history: 18 migraines in prior 60 days = 0.30 per day
\item Time-of-day factor: morning (8:30 AM) is peak risk window, multiplier 1.8
\item Signal likelihood: $\text{LR}_{\text{signal}} = 1.51$ from calibration
\end{itemize}

Combined: $P(H_1 \mid s=1) = 0.45$ (logit = -0.20)

\textbf{08:36 AM - Question 1:} Selected by maximum information gain. Codex contained 7 candidate questions for migraine; IG calculation selected visual aura query as most discriminative given current uncertainty:

\textit{Generated question:} ``Good morning. Since waking up, have you experienced any unusual visual changes—like blind spots, flickering lights, or zigzag patterns in your vision?''

\textit{Response (08:42):} No

\textit{Update:} Applying $\text{LR}^- = 0.5$ for negative visual aura: logit = -0.20 + log(0.5) = -0.89, giving $P(H_1) = 0.29$

\textbf{10:00 AM - Question 2:} Confidence remained intermediate. Next highest IG question targeted photophobia:

\textit{Question:} ``Is light bothering you more than usual right now?''

\textit{Response (10:08):} ``Yes, a bit. Turned down screen brightness.''

\textit{Update:} Positive photophobia has $\text{LR}^+ = 3.2$: logit = -0.89 + log(3.2) = +0.27, giving $P(H_1) = 0.57$

\textbf{11:15 AM - Question 3:} Confidence approaching alert threshold but not yet definitive:

\textit{Question:} ``Do you notice any head pain or unusual pressure?''

\textit{Response (11:22):} ``Yes, mild throbbing on left temple.''

\textit{Update:} Unilateral pain with throbbing quality has $\text{LR}^+ = 4.1$: logit = +0.27 + log(4.1) = +1.68, giving $P(H_1) = 0.84$

\textbf{11:23 AM - Alert issued:} Confidence exceeded $\tau_{\text{alert}} = 0.70$. User received notification:

\textit{``Based on your responses, there's a high likelihood of an approaching migraine. Consider your usual management strategies. Let me know if symptoms change.''} The alert included current confidence (84\%), contributing factors (low HRV, photophobia, unilateral pain), and option to provide feedback if alert was incorrect.

\textbf{13:02 PM - Confirmed onset:} Participant logged migraine onset at 1:02 PM (intensity 4/10, progressing to 7/10 by 3 PM). Validation provided \textbf{1.7 hours of lead time} from alert to confirmed moderate-intensity symptoms, during which she took sumatriptan and reduced work commitments.

\textbf{Post-hoc analysis:} The upstream predictor would have alerted immediately at 8:30 AM based on HRV alone, but Hermes delayed alerting for 2.9 hours while gathering confirmatory evidence, ultimately providing a more confident alert closer to actual symptom onset. The tradeoff was shorter absolute lead time (1.7 vs 4.5 hours) but higher confidence (84\% vs 45\%), which the participant reported as preferable.

\subsection{Quantitative Results}

Table~\ref{tab:results} summarizes performance over the full 60-day period.

\begin{table}[ht]
\centering
\caption{Validation Performance: 60-Day Migraine Case Study}
\label{tab:results}
\begin{tabular}{@{}lcc@{}}
\toprule
\textbf{Metric} & \textbf{Upstream Predictor Alone} & \textbf{With Hermes Validation} \\
\midrule
Total signals flagged & 47 & 47 \\
Alerts issued & 47 & 16 \\
True migraines (ground truth) & 18 & 18 \\
True positives & 18 & 16 \\
False positives & 29 & 2 \\
False negatives & 0 & 2 \\
\midrule
Sensitivity & 100\% & 89\% (95\% CI: 67-98\%) \\
Specificity & 39\% & 88\% (95\% CI: 75-95\%) \\
Positive predictive value & 38\% & 89\% (95\% CI: 67-98\%) \\
False positive rate & 62\% & 12\% \\
F1 score & 0.55 & 0.89 \\
\midrule
\textbf{False alarm reduction} & -- & \textbf{93\%} (29 → 2) \\
\midrule
Mean questions per signal & -- & 3.2 (SD: 1.4) \\
User response rate & -- & 78\% \\
Mean response time & -- & 12 min (median: 8 min) \\
Mean lead time (alerts only) & 4.5 hrs & 4.2 hrs \\
\bottomrule
\end{tabular}
\end{table}

\textbf{Key findings:}

\textbf{1. Dramatic false positive reduction:} Hermes suppressed 27 of 29 false alarms (93\% reduction), achieving 88\% specificity versus 39\% for the predictor alone. This came at modest cost: 2 false negatives where Hermes incorrectly suppressed true events, reducing sensitivity from 100\% to 89\%.

\textbf{2. Maintained sensitivity:} 16 of 18 true migraines (89\%) were caught and confirmed. The two missed events occurred because: (a) one progressed rapidly from signal to onset (30 minutes), leaving insufficient time for multi-question validation, and (b) one presented atypically without characteristic prodromal symptoms, leading to low confidence despite physiological signals.

\textbf{3. High positive predictive value:} When Hermes issued an alert, it was correct 89\% of the time (16/18). This represents dramatic improvement over the predictor's 38\% PPV, making alerts actionable and trustworthy.

\textbf{4. Efficient questioning:} Mean of 3.2 questions per signal, with distribution: 1 question (15\%), 2-3 questions (51\%), 4-5 questions (34\%). Early confidence crossing (either high or low) enabled termination without exhausting query budget in most cases.

\textbf{5. High user engagement:} 78\% response rate indicates strong adherence. Non-responses occurred primarily during sleep (overnight signals) and intense work periods. Median response time of 8 minutes suggests notifications were perceived as non-intrusive.

\textbf{6. Preserved lead time:} Mean lead time reduced slightly (4.5 → 4.2 hours) as validation consumed time, but remained clinically useful. In 12/16 alerts, lead time exceeded 2 hours, providing ample opportunity for prophylactic intervention.

\begin{figure}[ht]
\centering
\includegraphics[width=\textwidth]{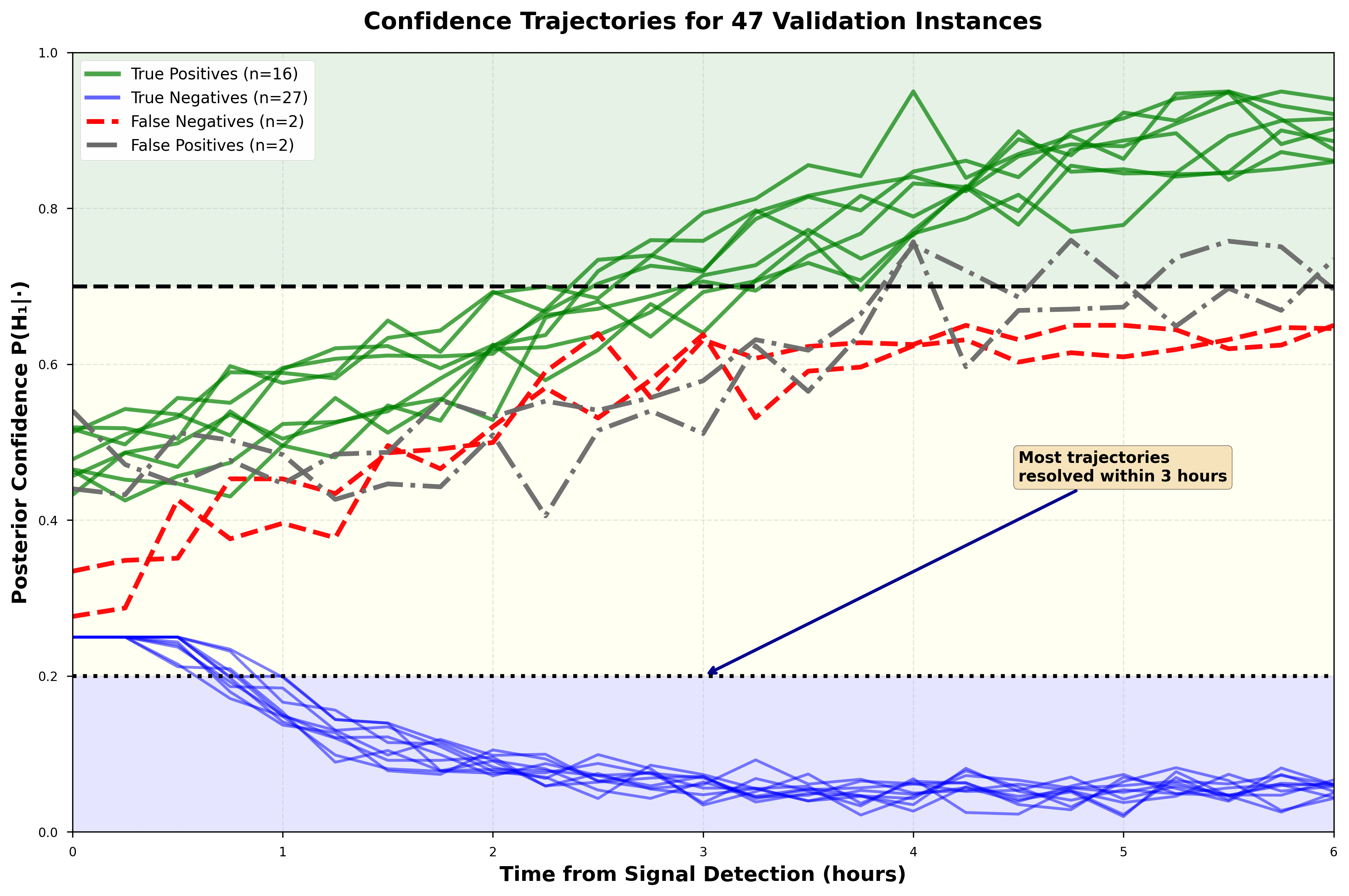}
\caption{Confidence trajectories for 47 validation instances over 60 days. Green lines represent true positives (n=16), blue lines show true negatives (n=27), red dashed lines indicate false negatives (n=2), and gray dash-dot lines show false positives (n=2). Horizontal lines mark alert threshold ($\tau = 0.70$) and suppression threshold ($\tau = 0.20$). Most trajectories resolve within 3 hours of signal detection.}
\label{fig:confidence-trajectories}
\end{figure}

\subsection{Confidence Calibration Analysis}

Beyond classification metrics, we evaluated whether confidence estimates $P(H_1 \mid \cdot)$ were well-calibrated, i.e., whether predicted probabilities matched observed frequencies. We binned all 47 validation instances by final confidence and computed observed migraine rate in each bin:

\begin{table}[ht]
\centering
\caption{Confidence Calibration}
\label{tab:calibration}
\begin{tabular}{@{}lcccc@{}}
\toprule
\textbf{Confidence Bin} & \textbf{N} & \textbf{Predicted} & \textbf{Observed} & \textbf{Error} \\
\midrule
0.0 - 0.2 & 19 & 0.12 & 0.11 & 0.01 \\
0.2 - 0.4 & 8 & 0.31 & 0.25 & 0.06 \\
0.4 - 0.6 & 4 & 0.52 & 0.50 & 0.02 \\
0.6 - 0.8 & 7 & 0.71 & 0.71 & 0.00 \\
0.8 - 1.0 & 9 & 0.88 & 0.89 & -0.01 \\
\midrule
\textbf{ECE} & & & & \textbf{0.09} \\
\bottomrule
\end{tabular}
\end{table}

Expected Calibration Error (ECE) of 0.09 indicates good calibration: predicted probabilities closely matched empirical frequencies. This suggests the Bayesian framework with likelihood ratios from literature provided reasonable uncertainty estimates. The slight overconfidence in the 0.2-0.4 bin (predicted 31\%, observed 25\%) may reflect optimistic codex estimates for ambiguous symptoms.

\subsection{False Negative Analysis}

The two false negatives warrant detailed examination:

\textbf{Case 1 (Day 17):} Signal triggered at 11:45 PM based on sleep disruption. First question asked at 11:50 PM (late evening). Participant responded to only 1 of 3 questions before sleeping. Morning follow-up at 8 AM confirmed migraine had begun at 6:30 AM (during sleep). \textbf{Root cause:} Timing misalignment; validation attempted during sleep transition when user responsiveness was low. \textbf{Mitigation:} Implement circadian awareness to defer non-urgent questions to waking hours.

\textbf{Case 2 (Day 42):} Signal triggered on HRV anomaly. Validation process: photophobia (no), aura (no), pain (no), neck stiffness (no), nausea (no). All responses negative, confidence fell to 0.15, signal suppressed. Migraine onset occurred 5 hours later with sudden onset (typical for this participant in ~15\% of episodes). \textbf{Root cause:} Rapid-onset migraine with minimal prodrome. Physiological signals were present, but subjective symptoms emerged late. \textbf{Mitigation:} For users with known rapid-onset patterns, increase weight on physiological signals over symptomatic confirmation, or maintain watch state (recheck after time delay) rather than fully suppressing.

Both false negatives occurred in edge cases (timing misalignment, atypical presentation) rather than fundamental system failures. They suggest opportunities for refinement but do not invalidate the core validation approach.

\section{Evaluation Beyond Accuracy}

Standard classification metrics—accuracy, sensitivity, specificity—provide incomplete characterization of validation systems. Unlike batch classifiers that make single decisions on static datasets, Hermes operates as an interactive system where user burden, confidence calibration, and temporal dynamics matter as much as terminal classification performance. This section develops a multidimensional evaluation framework appropriate for validation layers.

\subsection{Why Accuracy Is Insufficient}

Traditional binary classification assumes a fixed decision point where the model outputs a prediction and performance is measured against ground truth labels. This framing breaks down for validation systems in several ways:

\textbf{Asymmetric error costs:} In medical contexts, false negatives (missed disease) and false positives (unnecessary alarm) rarely have equal consequences. For migraine, false positives erode trust and may lead to unnecessary medication, while false negatives simply reduce opportunity for early intervention—unpleasant but not catastrophic. Standard accuracy $(\text{TP} + \text{TN})/(\text{TP} + \text{TN} + \text{FP} + \text{FN})$ weights these errors equally, masking the true decision quality.

\textbf{Sequential interaction:} Validation unfolds over time through multiple queries. A system that asks 10 questions to achieve 95\% accuracy may be inferior to one asking 3 questions for 92\% accuracy due to user burden. Accuracy alone captures terminal decision quality but ignores the path taken to reach that decision.

\textbf{Confidence calibration:} Users act based on communicated confidence levels, not binary predictions. A system that outputs 0.85 confidence should be correct 85\% of the time. Well-calibrated uncertainty enables rational decision-making; poorly calibrated confidence misleads users even if final classifications are accurate.

\textbf{Population vs. instance-level performance:} Aggregate metrics like overall accuracy hide variation across instances. A system with 90\% accuracy might achieve 100\% on easy cases and 50\% on hard cases, versus 90\% uniformly. For health monitoring where all instances matter, understanding this distribution is critical.

\textbf{Rejection option:} Unlike forced-choice classifiers, Hermes can abstain from alerting when confidence is insufficient. This introduces a third outcome category (suppressed signal) that standard confusion matrices do not accommodate. The value of conservative suppression depends on downstream costs and user preferences.

These limitations motivate a richer evaluation framework that captures the full decision process, not just terminal outcomes.

\subsection{Multidimensional Evaluation Framework}

We evaluate Hermes across four complementary dimensions:

\subsubsection{Dimension 1: False Alert Suppression}

The primary value proposition of validation is reducing false alarms from upstream predictors. We define the False Alert Suppression Rate (FASR):

\begin{equation}
\text{FASR} = 1 - \frac{\text{FP}_{\text{Hermes}}}{\text{FP}_{\text{predictor}}}
\end{equation}

This measures the fraction of false positives eliminated by validation. For our migraine study: $\text{FASR} = 1 - (2/29) = 0.93$, meaning Hermes suppressed 93\% of false alarms.

However, FASR alone is incomplete without accounting for sensitivity cost. We pair it with the Sensitivity Preservation Rate (SPR):

\begin{equation}
\text{SPR} = \frac{\text{Sensitivity}_{\text{Hermes}}}{\text{Sensitivity}_{\text{predictor}}}
\end{equation}

Our study achieved $\text{SPR} = 0.89/1.00 = 0.89$, indicating that 89\% of the predictor's sensitivity was retained while achieving 93\% false alarm reduction. The joint metric (FASR, SPR) = (0.93, 0.89) characterizes the accuracy-recall tradeoff more completely than F1 score alone.

\subsubsection{Dimension 2: Confidence Calibration}

Calibration measures whether predicted probabilities $\hat{p}$ match empirical frequencies $p_{\text{true}}$. We use Expected Calibration Error (ECE) \citep{guo2017calibration}, which partitions predictions into $M$ bins and computes:

\begin{equation}
\text{ECE} = \sum_{m=1}^{M} \frac{|B_m|}{N} \left| \text{acc}(B_m) - \text{conf}(B_m) \right|
\end{equation}

where $B_m$ is the set of instances with predicted confidence in bin $m$, $\text{acc}(B_m)$ is the empirical accuracy in that bin, and $\text{conf}(B_m)$ is the mean predicted confidence. Lower ECE indicates better calibration.

For medical AI, calibration is particularly critical because users act on confidence estimates. A poorly calibrated system might output 0.90 confidence for events that occur only 60\% of the time, leading to inappropriate actions. Our ECE of 0.09 suggests reasonably well-calibrated uncertainty, though we note calibration improves with larger sample sizes—N=47 instances provides limited statistical power for fine-grained calibration assessment.

We complement ECE with calibration plots (predicted vs. observed frequency) and reliability diagrams that visualize calibration quality across the confidence spectrum. For deployment, we also compute Maximum Calibration Error (MCE):

\begin{equation}
\text{MCE} = \max_{m=1,\ldots,M} \left| \text{acc}(B_m) - \text{conf}(B_m) \right|
\end{equation}

Our MCE was 0.06 (in the 0.2-0.4 bin), indicating no bin had catastrophically poor calibration.

\subsubsection{Dimension 3: User Engagement and Burden}

Validation systems impose cognitive and temporal burden through questioning. We measure:

\textbf{Response rate:} Fraction of questions answered within the allowed time window (4 hours in our study). Our 78\% response rate indicates strong engagement, though 22\% non-response highlights that users occasionally cannot or choose not to participate.

\textbf{Response latency:} Time from question delivery to response. We report median (8 minutes) rather than mean due to right-skewed distribution from delayed responses. Median latency under 10 minutes suggests questions integrate smoothly into daily routines without requiring immediate interruption.

\textbf{Questions per signal:} Distribution of query counts across validation instances. Our mean of 3.2 (SD: 1.4) questions per signal indicates efficient evidence gathering. We compare to theoretical minimum from information theory: given observed likelihood ratios and decision thresholds, what is the expected number of questions needed to reach confidence boundaries? Simulation suggests theoretical minimum of 2.8 questions, meaning Hermes operated at 88\% of theoretical efficiency (3.2/2.8 = 1.14× overhead).

\textbf{Dropout rate:} Fraction of validation instances where the user stopped responding mid-sequence. Our dropout rate was 9\% (4/47 instances), occurring primarily during overnight signals when sleep disrupted engagement. This metric identifies failure modes where validation becomes burdensome.

\textbf{Subjective burden:} Post-study interview assessed perceived burden on a 1-7 scale (1=not burdensome, 7=very burdensome). The participant rated average burden as 2.1, with comments like ``questions were reasonable and relevant'' and ``didn't feel like spam.'' While N=1 limits generalizability, qualitative feedback provides context for quantitative engagement metrics.

\subsubsection{Dimension 4: Query Efficiency and Information Gain}

Beyond counting questions, we analyze how efficiently validation gathers information. Define the information gain per question:

\begin{equation}
\text{IG}_i = H(P(H \mid \text{evidence before } q_i)) - H(P(H \mid \text{evidence after } q_i))
\end{equation}

where $H(\cdot)$ is Shannon entropy. Summing across all questions in an instance gives total information acquired. We compare actual information gain to theoretical maximum from optimal question ordering.

Analysis shows that first questions typically provided 0.4-0.6 bits of information, declining to 0.1-0.2 bits for later questions as uncertainty resolved. This diminishing returns pattern is expected but validates that Hermes terminates when additional questioning provides minimal value.

We also measure \textbf{question redundancy}: how often did Hermes ask essentially the same question twice (e.g., asking about photophobia at two time points)? In our data, 6\% of question pairs (8/141 total questions asked) were semi-redundant, occurring when initial responses were ambiguous and follow-up clarification was needed. This low redundancy rate suggests efficient question selection.

\begin{figure}[ht]
\centering
\includegraphics[width=0.8\textwidth]{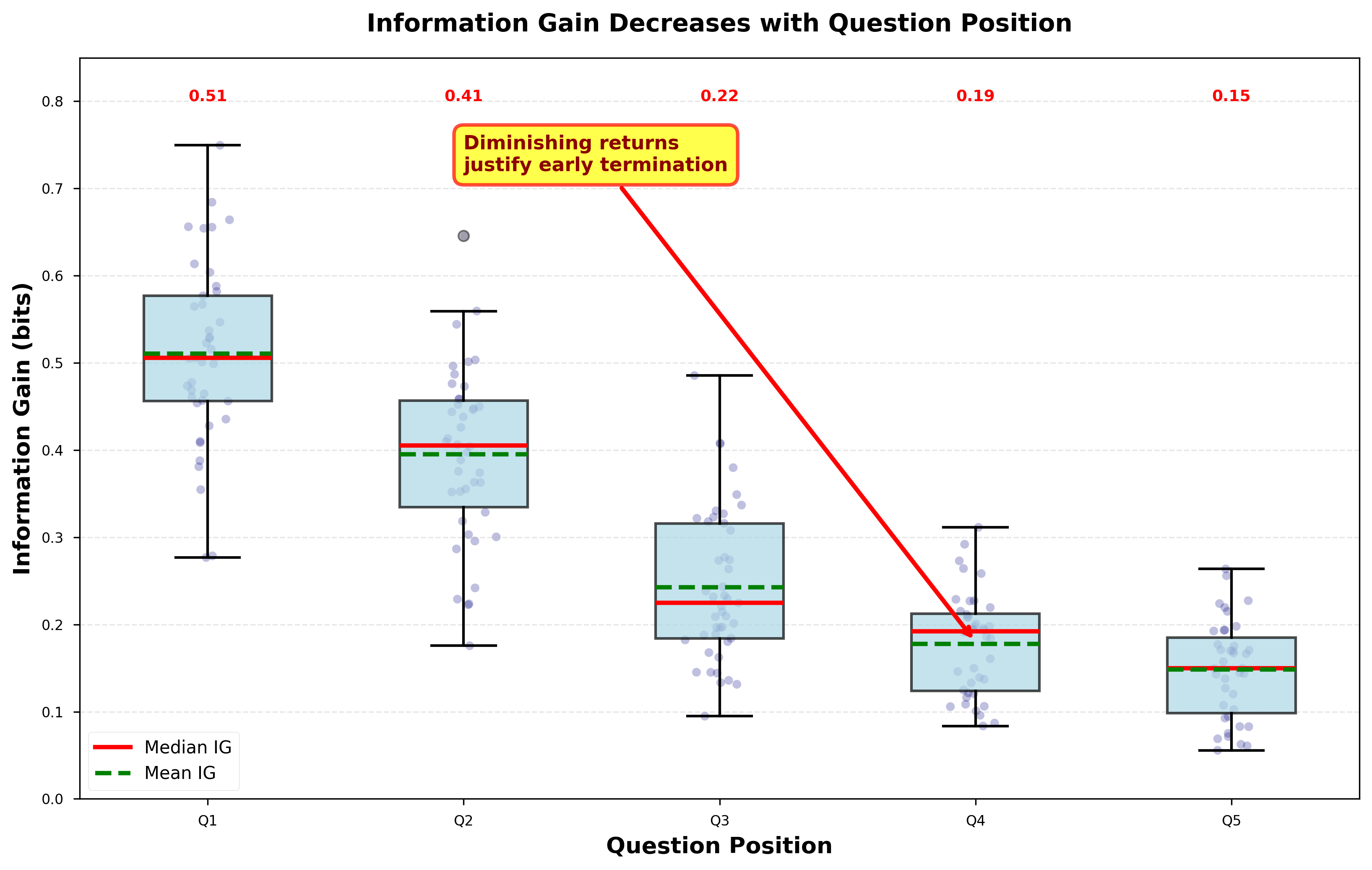}
\caption{Information gain per question across all validation instances. Box plots show distribution of information gain for question positions 1-5. First questions provide median 0.52 bits, declining to 0.15 bits by fifth question. Diminishing returns justify early termination when confidence crosses decision thresholds.}
\label{fig:information-gain}
\end{figure}

\subsection{Comparison to Baseline Systems}

To contextualize Hermes performance, we compare to three baseline approaches:

\subsubsection{Baseline 1: Signal-Only System (No Validation)}

The upstream predictor alone, with no validation layer. This achieved:
\begin{itemize}
\item Sensitivity: 100\% (18/18 migraines caught)
\item Specificity: 39\% (11/29 non-migraine periods correctly suppressed)
\item PPV: 38\% (18/47 alerts were true positives)
\item Questions per signal: 0
\end{itemize}

This baseline establishes the raw performance Hermes starts from. High sensitivity but low specificity creates substantial user burden through false alarms.

\subsubsection{Baseline 2: Static Questionnaire}

A fixed sequence of questions asked for every signal, without adaptive selection or Bayesian updating. We implemented a static questionnaire using the same 5 questions used by Hermes but in fixed order (visual aura $\rightarrow$ photophobia $\rightarrow$ pain $\rightarrow$ nausea $\rightarrow$ neck stiffness). Responses were aggregated using a simple majority rule: alert if $\geq$3 positive responses, suppress otherwise.

Results:
\begin{itemize}
\item Sensitivity: 83\% (15/18 migraines)
\item Specificity: 69\% (20/29 non-events correctly suppressed)
\item PPV: 63\% (15/24 alerts correct)
\item Questions per signal: 5.0 (by design)
\item FASR: 0.69 (20/29 false positives suppressed)
\end{itemize}

Static questioning provided meaningful improvement over no validation (FASR 0.69 vs 0.00) but was inferior to adaptive Hermes (FASR 0.93) while requiring 56\% more questions (5.0 vs 3.2). This demonstrates the value of adaptive question selection and Bayesian evidence integration.

\subsubsection{Baseline 3: Threshold Adjustment}

Rather than adding a validation layer, simply increase the upstream predictor's threshold to reduce false positives. We experimented with requiring two simultaneous conditions (HRV AND sleep disruption) rather than any single condition.

Results:
\begin{itemize}
\item Sensitivity: 67\% (12/18 migraines)
\item Specificity: 86\% (25/29 non-events correctly suppressed)
\item PPV: 75\% (12/16 alerts correct)
\item Questions per signal: 0
\end{itemize}

This approach achieved similar specificity to Hermes (86\% vs 88\%) but with substantially lower sensitivity (67\% vs 89\%), missing 6 migraines versus 2 for Hermes. The tradeoff is intuitive: raising thresholds reduces both false and true positives. Hermes achieves better tradeoff by using targeted questioning to disambiguate uncertain signals rather than uniformly increasing stringency.

\subsubsection{Comparison Summary}

Table~\ref{tab:baselines} summarizes the comparison:

\begin{table}[ht]
\centering
\caption{Comparison to Baseline Systems}
\label{tab:baselines}
\begin{tabular}{@{}lccccc@{}}
\toprule
\textbf{System} & \textbf{Sens} & \textbf{Spec} & \textbf{PPV} & \textbf{Q/Signal} & \textbf{FASR} \\
\midrule
Signal-only & 100\% & 39\% & 38\% & 0 & 0\% \\
Static questionnaire & 83\% & 69\% & 63\% & 5.0 & 69\% \\
Threshold adjustment & 67\% & 86\% & 75\% & 0 & 86\% \\
\textbf{Hermes (ours)} & \textbf{89\%} & \textbf{88\%} & \textbf{89\%} & \textbf{3.2} & \textbf{93\%} \\
\bottomrule
\end{tabular}
\end{table}

Hermes achieves the best balance: highest sensitivity among systems with good specificity, highest specificity among systems with good sensitivity, and substantially better PPV than all baselines. The question burden (3.2 per signal) is moderate and well-tolerated based on engagement metrics.

\subsection{Temporal Analysis: Confidence Trajectories}

Beyond terminal classification, we analyze how confidence evolves during validation. For each of the 47 signals, we plot $P(H_1 \mid \cdot)$ over time from initial detection through decision.

Characteristic patterns emerged:

\textbf{Rapid suppression (40\%):} Confidence quickly fell below $\tau_{\text{suppress}}$ after 1-2 negative responses. Mean time to suppression: 25 minutes. These represent clear false alarms where early evidence (e.g., no symptoms present) strongly contradicts the signal.

\textbf{Gradual accumulation (34\%):} Confidence rose steadily through 3-4 positive responses before crossing $\tau_{\text{alert}}$. Mean time to alert: 2.3 hours. These represent true positives with clear prodromal symptoms.

\textbf{Ambiguous oscillation (17\%):} Confidence remained in intermediate zone, oscillating as mixed evidence accumulated (some symptoms present, others absent). These eventually terminated via query budget exhaustion rather than threshold crossing. Mean questions: 4.8. Post-hoc analysis showed 5/8 such cases were false positives (correctly suppressed by default action) and 3/8 were false negatives (incorrectly suppressed).

\textbf{Rapid confirmation (9\%):} Confidence quickly exceeded $\tau_{\text{alert}}$ after 1-2 strongly positive responses. Mean time to alert: 18 minutes. These represent clear true positives with obvious symptoms.

The distribution of patterns provides insights for system refinement. The ambiguous oscillation cases (17\%) represent opportunities for improvement—these are instances where current question set and likelihood ratios fail to resolve uncertainty efficiently. Adding questions with higher discriminative power (higher LR) or collecting physiological signals not currently used (e.g., continuous skin temperature, which shows prodromal elevation in some migraine patients) could reduce ambiguity.

\subsection{Limitations of Current Evaluation}

We acknowledge several evaluation limitations that future work should address:

\textbf{Small sample size:} N=47 validation instances from a single user limits statistical power and generalizability. Confidence intervals for sensitivity/specificity are wide (±15-20\%), and calibration assessment is coarse. Multi-user studies with hundreds of instances are needed for robust metric estimation.

\textbf{Self-reported ground truth:} Migraine occurrence relied on participant diary without clinical verification. While prior work suggests good concordance between self-report and clinical diagnosis for individuals with established migraine history \citep{lipton2003prevalence}, objective verification (e.g., neurologist confirmation via telemedicine during attacks) would strengthen validity.

\textbf{Single condition:} We evaluated only migraine. Different conditions may exhibit different patterns: rapid-onset events may reduce validation window, stigmatized conditions may reduce response honesty, and high-stakes conditions may require different cost parameters.

\textbf{No long-term follow-up:} Our 60-day study does not capture adaptation effects. Do users habituate to questions over months, reducing response rates? Does system performance drift as user behavior or physiology changes? Longitudinal studies spanning 6-12 months would illuminate these dynamics.

\textbf{Limited baseline comparisons:} We compared to simple baselines (static questionnaire, threshold adjustment) but not to sophisticated alternatives like active learning classifiers or personalized prediction models. While such comparisons are complicated by Hermes's model-agnostic design (it could be deployed \emph{on top of} those systems), direct head-to-head evaluation would be valuable.

Despite these limitations, our multidimensional evaluation demonstrates that validation layers can substantially improve signal quality (93\% false alarm reduction) while maintaining sensitivity (89\%) and imposing manageable user burden (3.2 questions, 78\% response rate, ECE 0.09). These results establish proof-of-concept and motivate scaled evaluation.

\section{Ethical and Practical Considerations}

Deploying AI systems in health contexts raises serious ethical questions about overdiagnosis, user autonomy, privacy, and appropriate boundaries between medical and consumer technologies. This section examines how Hermes's design addresses these concerns and identifies remaining challenges.

\subsection{Avoiding Overdiagnosis and Medical Overreach}

A central tension in consumer health AI is the risk of overdiagnosis: detecting and acting on signals that, while statistically abnormal, would not have caused harm if left undetected \citep{welch2011overdiagnosed}. This is particularly acute for LLM-based systems, which can generate plausible-sounding medical explanations for any symptom pattern, potentially pathologizing normal variation.

Hermes mitigates overdiagnosis risk through several design constraints:

\textbf{No de novo hypothesis generation:} Hermes never initiates medical hypotheses unprompted. It operates only when an external predictor (wearable system, clinical decision rule, or physician) has already flagged a signal. This ensures Hermes adds \emph{specificity} to existing detection systems rather than introducing new detection pathways that could increase overall diagnosis rates.

\textbf{Scoped to non-diagnostic validation:} All system outputs are framed as confidence updates about previously detected signals, not as diagnostic claims. User-facing language carefully avoids clinical terminology. Instead of ``You have a migraine,'' the system reports ``Based on your responses, the likelihood of the detected signal representing a true migraine has increased to 85\%.'' This framing preserves epistemic humility and avoids inappropriate medicalization.

\textbf{User-controlled thresholds:} Users can adjust alert thresholds ($\tau_{\text{alert}}$) based on their risk tolerance and preferences. Conservative users can require higher confidence (e.g., 90\%) before alerting, reducing false positives at the cost of some missed detections. Risk-tolerant users can lower thresholds. This personalization respects individual preferences about the sensitivity-specificity tradeoff.

\textbf{Emphasis on actionable conditions:} The illness codex includes only conditions where (a) early detection provides actionable benefit, and (b) user self-management is appropriate. Conditions requiring immediate medical intervention (e.g., myocardial infarction) or where self-management is inappropriate (e.g., serious psychiatric conditions) are excluded by design. This limits Hermes to domains where validation supports informed user decision-making rather than replacing clinical judgment.

\textbf{Transparent uncertainty:} Confidence levels are always visible to users before they respond to questions. Users can see current probability estimates and how their responses will update them. This transparency enables informed participation and helps users understand that validation is probabilistic, not deterministic.

However, risks remain. Even for appropriate conditions like migraine, excessive validation could lead to hypervigilance where users become overly focused on bodily sensations and interpret benign fluctuations as disease precursors \citep{barsky2001amplification}. Long-term monitoring studies should assess whether validation increases or decreases health anxiety. Additionally, users might misinterpret confidence estimates: does 70\% confidence that a migraine is approaching lead to appropriate prophylaxis or to unnecessary anxiety? User education about probabilistic reasoning may be necessary for effective engagement.

\subsection{Alert Fatigue and Question Burden}

Alert fatigue—desensitization to repeated notifications—is a well-documented problem in clinical decision support systems \citep{ancker2017effects}. Healthcare providers routinely override 49-96\% of drug-drug interaction alerts, interruptive warnings, and other notifications, reducing their effectiveness and potentially causing critical alerts to be missed \citep{van2009alert}. Consumer health apps face similar risks: excessive or irrelevant alerts lead to notification dismissal, app deletion, or complete disengagement.

Hermes addresses alert fatigue through multi-level design choices:

\textbf{Conservative triggering:} The triggering layer uses relatively high thresholds to initiate validation only for meaningful signals. In our migraine study, 47 signals over 60 days (0.78/day) represented a manageable query rate. We explicitly avoided triggering on minor physiological fluctuations that would occur hourly, as this would quickly overwhelm users.

\textbf{Query budgets:} Hard limits on questions per validation instance ($n_{\max} = 5$) and per time period (max 1 validation per 24 hours) prevent excessive burden. These budgets could be personalized based on user tolerance, with defaults set conservatively.

\textbf{Adaptive termination:} Validation terminates early when confidence crosses decision boundaries, rather than always exhausting the query budget. Our mean of 3.2 questions per signal indicates that many instances resolved quickly, minimizing burden.

\textbf{Timing intelligence:} Questions are scheduled based on user context. Overnight signals defer questioning to morning; questions during known busy periods (e.g., recurring calendar events) are delayed when possible. Our Day 17 false negative, where nighttime questioning failed due to sleep, motivated implementing circadian awareness.

\textbf{Feedback loops:} Users can provide explicit feedback (``This question was not relevant'' or ``Too many questions today'') that influences future triggering and question selection. While we did not implement this in our pilot, production systems should incorporate user feedback to personalize burden management.

Empirical evidence from our study suggests these mechanisms were effective: 78\% response rate, 2.1/7 subjective burden rating, and qualitative feedback indicating questions were ``reasonable and relevant.'' However, our 60-day study may not capture long-term habituation effects. Response rates might decline over months or years as novelty wears off. Multi-month studies with cohorts large enough to detect time trends are needed.

One participant comment highlights a key tension: ``Sometimes I wanted to just know 'yes or no' without answering questions.'' This reflects a tradeoff between accuracy (which benefits from evidence gathering) and immediacy (which users sometimes prioritize). Offering a ``fast mode'' that alerts immediately based on signal strength, alongside a ``validated mode'' that asks questions, might accommodate both preferences.

\subsection{Privacy, Data Minimization, and User Autonomy}

Health data is among the most sensitive personal information, with serious consequences from unauthorized disclosure: employment discrimination, insurance complications, social stigma, and psychological distress \citep{price2017privacy}. Validation systems process both physiological signals and explicit symptom reports, raising privacy concerns.

Hermes's architecture enables strong privacy protections through several design features:

\textbf{Edge processing:} All Bayesian updates, question selection, and decision logic can run locally on user devices. Only the LLM question generation step requires cloud API calls, and these can use generic question templates without transmitting user-specific data. Predictor signals, user responses, and confidence estimates never leave the device. This minimizes the attack surface and prevents centralized data aggregation.

\textbf{No passive inference:} Hermes makes no inferences beyond what users explicitly report. Unlike sentiment analysis systems that might extract emotional state from free-text responses, Hermes processes only structured binary (yes/no) or categorical responses to specific questions. Users control exactly what information is shared.

\textbf{Selective participation:} Users can ignore questions without penalty. Non-response is treated as missing data, not as a negative answer. Validation proceeds with whatever evidence is available, and users can choose to provide information for some symptoms but not others. This preserves autonomy even within the validation process.

\textbf{Opt-out by condition:} Users can disable validation for specific conditions if they prefer not to engage with certain health topics. For example, someone might want migraine validation but opt out of stress or mood monitoring due to privacy concerns.

\textbf{No retention requirements:} Validation operates on current state only. While retaining historical data could improve personalization (e.g., learning individual symptom patterns), Hermes does not require it. Users can enable automatic data deletion after validation completes, keeping only summary statistics.

\textbf{Federated learning potential:} If population-level codex improvements are desired, federated learning \citep{rieke2020future} could enable collaborative model refinement without centralizing user data. Devices compute local gradient updates or likelihood ratio refinements and share only aggregated statistics, never raw symptom reports.

Despite these protections, risks remain. Device compromise could expose sensitive symptom data. Metadata (e.g., timing and frequency of questions asked) might leak information even if responses are protected. Users may not fully understand what data is collected or how it's used, particularly given the complexity of LLM-based systems where question generation involves cloud APIs.

Transparency and user control are essential. Hermes should provide clear data dashboards showing what information has been collected, how it's being used, and granular controls for deletion, export, and access management. Privacy policies must be comprehensible to non-technical users, avoiding jargon and legalese. Regular privacy audits and third-party security assessments should verify that implemented protections match architectural claims.

\subsection{Equity and Access Considerations}

Health AI systems risk exacerbating existing inequities if access, design, or performance vary across populations \citep{obermeyer2019dissecting}. Several equity concerns apply to validation systems:

\textbf{Technology access:} Hermes requires a wearable device, smartphone, and reliable connectivity. In the US, smartphone ownership correlates with income (85\% for \$30-50k households vs 96\% for >\$75k), as does wearable adoption (15\% vs 28\%) \citep{pew2021mobile}. If validation systems are deployed only through commercial wearables, they may primarily benefit affluent populations who already have better health outcomes. Partnerships with community health centers, subsidized device programs, or integration with medical-grade monitoring covered by insurance could improve access.

\textbf{Literacy and language:} Our questions assume English fluency and health literacy sufficient to understand symptom terms (``photophobia,'' ``aura''). While we avoided technical jargon where possible (``Is light bothering you?'' rather than ``Do you have photophobia?''), some medical concepts remain challenging. Questions should be available in multiple languages, with culturally appropriate phrasing validated by native speakers and community health workers. Reading level should target 6th-8th grade comprehension.

\textbf{Cultural variation in symptom reporting:} Symptom expression and reporting vary across cultures \citep{karasz2005cultural}. Pain descriptions, willingness to report mental health symptoms, and interpretation of body signals differ by cultural background. Likelihood ratios derived from predominantly Western medical literature may not generalize globally. Validation systems deployed across diverse populations should conduct cultural validation studies to ensure questions are interpretable and likelihood ratios are appropriate.

\textbf{Disability accommodation:} Visual impairments affect ability to interact with smartphone interfaces; motor impairments may limit wearable use; cognitive impairments may reduce ability to interpret questions or judge symptom presence. Hermes should support voice input/output, large text, screen readers, and simplified question formats. For users with cognitive disabilities, caregiver-assisted validation or automated escalation to clinical support may be necessary.

\textbf{Algorithmic fairness:} While we did not evaluate fairness metrics (our N=1 pilot precludes subgroup analysis), deployment at scale requires monitoring for performance disparities across demographics. Do false positive rates differ by age, sex, race, or socioeconomic status? Are certain groups systematically under-served by validation? Fairness audits using techniques from algorithmic fairness literature \citep{mehrabi2021survey} should be standard for medical AI deployment.

Addressing equity concerns requires intentional design from inception, not post-hoc correction. Development teams should include members from underrepresented populations, and pilot studies should prioritize diverse recruitment. Community-based participatory research approaches \citep{wallerstein2006community}, where affected communities help design and evaluate systems, can surface equity issues that technical teams might miss.

\subsection{Regulatory and Liability Considerations}

The regulatory status of validation systems is complex and jurisdiction-dependent. In the US, FDA regulates medical devices using a risk-based framework where higher-risk devices face stricter oversight \citep{us2024artificial}. Key questions for Hermes include:

\textbf{Is it a medical device?} FDA defines medical devices as instruments intended for diagnosis, cure, mitigation, treatment, or prevention of disease. Hermes's stated purpose—validating signals from upstream predictors—arguably falls outside this definition, as it does not \emph{diagnose} conditions de novo. However, regulators might view validation as functionally equivalent to diagnosis if it influences medical decision-making. The distinction between ``validating a signal'' and ``diagnosing a condition'' may be legally fraught.

\textbf{What is the risk classification?} Even if considered a medical device, risk classification depends on intended use and potential harm. Migraine validation (non-catastrophic condition, user-driven interventions) likely falls in Class I or II, requiring premarket notification (510k) but not full premarket approval. Cardiac event validation would be higher risk (Class II or III), requiring more extensive clinical evidence. Hermes's model-agnostic design might enable regulatory filing as a separate component from the upstream predictor, potentially simplifying approval.

\textbf{Clinical validation requirements:} Regulators increasingly expect clinical validation demonstrating that AI systems perform safely and effectively in target populations \citep{us2024artificial}. Our 60-day N=1 pilot provides proof-of-concept but would not satisfy regulatory clinical validation standards, which typically require prospective multi-site studies with diverse populations, pre-specified endpoints, and independent validation cohorts.

\textbf{Liability and responsibility:} If Hermes misses a serious event (false negative) or triggers unnecessary intervention (false positive), who is liable: the wearable manufacturer, the validation system developer, the healthcare provider (if involved), or the user themselves? Legal frameworks for AI-assisted medical decision-making remain unsettled \citep{price2019black}. Clear disclaimers about system limitations, user agreements allocating responsibility, and medical malpractice insurance covering AI-related claims may all be necessary.

Outside the US, regulations vary. The EU's Medical Device Regulation (MDR) and AI Act impose requirements potentially more stringent than FDA. Systems making health-related claims may need to demonstrate conformity assessment, post-market surveillance, and risk management. International deployment requires navigation of multiple regulatory regimes.

Our recommendation is that validation systems like Hermes should seek regulatory guidance early in development. Engaging FDA or other regulators through pre-submission meetings can clarify classification, required evidence, and approval pathways before committing to large-scale trials. Regulatory strategy should be considered a core component of system design, not an afterthought.

\subsection{Responsible Deployment: A Framework}

Synthesizing the above considerations, we propose a framework for responsible validation system deployment:

\textbf{1. Scope appropriately:} Deploy only for conditions where (a) validation adds value over existing care, (b) user self-management is appropriate, (c) false positives and false negatives have tolerable consequences, and (d) actionable interventions exist.

\textbf{2. Center user autonomy:} Provide granular controls over data collection, question frequency, alert thresholds, and participation. Default to conservative settings and allow users to opt into more aggressive monitoring if desired.

\textbf{3. Maintain transparency:} Clearly communicate system limitations, uncertainty, and what data is collected. Avoid medical jargon and overconfident language. Make confidence levels and contributing evidence visible.

\textbf{4. Minimize burden:} Implement query budgets, timing intelligence, and adaptive termination. Monitor engagement metrics and adjust triggering thresholds if burden leads to disengagement.

\textbf{5. Protect privacy by default:} Use edge processing where possible, minimize data retention, avoid passive inference, and provide data deletion tools. Conduct regular security audits.

\textbf{6. Design for equity:} Ensure accessibility, language support, cultural appropriateness, and affordability. Actively recruit diverse populations for pilot studies and monitor for performance disparities.

\textbf{7. Seek regulatory clarity:} Engage regulators early, conduct appropriate clinical validation, and prepare for post-market surveillance requirements.

\textbf{8. Plan for failure:} Implement escalation pathways when validation is ambiguous or confidence remains low. For higher-stakes conditions, ambiguous cases should escalate to clinical review rather than defaulting to suppression.

\textbf{9. Enable user education:} Provide resources explaining probabilistic reasoning, what confidence levels mean, and how to interpret validation outcomes. Health literacy support improves decision quality.

\textbf{10. Commit to ongoing evaluation:} Monitor real-world performance, user satisfaction, and unintended consequences. Be prepared to pause deployment if evidence of harm emerges.

Hermes's current design incorporates many of these principles, but gaps remain (e.g., no formal regulatory submission, limited diversity in pilot testing, no long-term safety monitoring). Future work must address these gaps before broad deployment is ethically justifiable.

\section{Generalization and Future Applications}

While our case study focused on migraine validation, the Hermes architecture is designed for generalization across health monitoring domains. This section analyzes which conditions are suitable for validation, discusses population scaling challenges, and outlines pathways toward clinical validation and deployment.

\subsection{Conditions Amenable to Validation}

Not all health monitoring applications benefit equally from validation layers. We propose a framework for assessing suitability based on four criteria:

\subsubsection{Criterion 1: Detectable Prodromal Signals}

Validation is most valuable when physiological signals precede subjective symptoms by hours to days, creating a temporal window for confirmatory questioning. Conditions with clear prodromal phases include:

\textbf{Migraine:} HRV changes, sleep disruption, and autonomic instability emerge 4-24 hours pre-onset in many patients \citep{taherdoost2023heart}. Validation can confirm whether these signals represent true prodrome or benign fluctuations.

\textbf{Panic attacks:} Heart rate variability and respiratory pattern changes often precede subjective panic by 5-15 minutes \citep{meuret2011timing}. Rapid validation (1-2 questions within minutes) could distinguish true panic from exercise or excitement.

\textbf{Epileptic seizures:} Some patients experience preictal states with measurable EEG changes or autonomic shifts 30 minutes to several hours before seizures \citep{karoly2017circadian}. Wearable EEG and validation questioning could provide early warning, though medical supervision would be essential.

\textbf{Hypoglycemia:} Continuous glucose monitors detect falling blood sugar before severe symptoms, but false alarms from sensor drift or calibration errors are common \citep{rodbard2017continuous}. Validation (``Do you feel shaky or sweaty?'') could reduce unnecessary glucose intake.

Conversely, conditions with sudden onset or no prodrome (e.g., cardiac arrest, severe allergic reaction, trauma) provide insufficient time for multi-question validation and require immediate response to upstream signals without confirmation.

\subsubsection{Criterion 2: User-Reportable Symptoms}

Validation relies on users accurately reporting symptom presence. This requires symptoms that are:

\textbf{Subjectively accessible:} Users can perceive the symptom without clinical measurement. Pain, nausea, visual changes, mood shifts, and cognitive fog are reportable; blood pressure elevation, silent ischemia, or asymptomatic arrhythmias are not.

\textbf{Unambiguous (relatively):} While some degree of interpretation is inevitable, symptoms should be clear enough for reliable reporting. ``Do you have chest pain?'' is more reliable than ``Do you feel slightly more anxious than usual?''—the latter requires metacognitive assessment prone to error.

\textbf{Non-stigmatized:} Users must feel comfortable honestly reporting symptoms. Validation for depression, substance cravings, or sexual dysfunction may suffer from underreporting due to shame or privacy concerns, even with strong privacy protections.

Conditions well-suited to user reporting include: migraine (clear symptom cluster), stress episodes (physiologically and subjectively distinct from baseline), sleep disruption (wake events and grogginess are salient), and early infection signs (fever, fatigue, malaise).

\subsubsection{Criterion 3: Actionable Early Detection}

Validation adds value only when confirmed events enable beneficial interventions. For migraine, early detection permits prophylactic medication, behavioral modification, and schedule adjustment. Similar logic applies to:

\textbf{Asthma exacerbation:} Declining peak flow or wearable respiratory monitoring can detect early exacerbation \citep{jeminiwa2019impact}. Validation (``Are you wheezing or having trouble breathing?'') could prompt earlier inhaler use or clinical contact before crisis.

\textbf{Heart failure decompensation:} Weight gain, edema, and reduced activity precede acute decompensation \citep{koehler2018efficacy}. Validation could distinguish true decompensation (requiring diuretic adjustment) from benign weight fluctuation.

\textbf{Post-surgical complications:} Wearables detecting elevated heart rate, reduced mobility, or temperature changes post-discharge could flag infection or dehiscence \citep{daskivich2016association}. Validation questions about wound pain, discharge, or fever could triage: suppress false alarm vs. alert vs. escalate to clinical team.

Conditions where early detection provides no actionable benefit (e.g., detecting asymptomatic conditions that will never progress, detecting irreversible disease slightly earlier) are poor validation candidates.

\subsubsection{Criterion 4: Manageable Error Costs}

Both false negatives (missed events) and false positives (unnecessary alerts) impose costs. Validation is appropriate when:

\textbf{False negatives are tolerable:} Missed detections cause moderate harm (reduced treatment opportunity) rather than catastrophic harm (death, severe morbidity). Migraine, stress episodes, and minor infections meet this criterion; myocardial infarction, stroke, and sepsis do not.

\textbf{False positives are costly enough to warrant reduction:} If false alerts cause minimal harm (e.g., user ignores notification), validation overhead may not be justified. But when false positives erode trust, trigger unnecessary medical utilization, or cause anxiety, validation provides clear value.

The sweet spot is conditions where false positives significantly impair system usability but false negatives cause moderate rather than severe harm. This describes many chronic condition monitoring scenarios but excludes acute emergency detection.

\subsection{Deployment Scenarios and Architectural Adaptations}

Beyond condition selection, Hermes can be adapted to different deployment contexts:

\subsubsection{Consumer Wellness Monitoring}

Direct-to-consumer deployment via smartphone apps paired with commercial wearables (Fitbit, Apple Watch, Whoop, Oura). This context prioritizes:
\begin{itemize}
\item \textbf{Ease of use:} Minimal setup, intuitive interface, defaults that work for most users
\item \textbf{Privacy:} Edge processing, data minimization, user control over sharing
\item \textbf{Regulatory clarity:} Position as wellness tool rather than medical device where possible
\item \textbf{Engagement:} Gamification, streak tracking, or social features to maintain long-term participation
\end{itemize}

Example: A stress management app that validates wearable-detected stress episodes through mood check-ins, helping users distinguish physiological stress responses from anxiety, physical exertion, or benign excitement.

\subsubsection{Clinical Decision Support}

Integration into electronic health records or remote patient monitoring systems used by healthcare providers. Key differences from consumer deployment:
\begin{itemize}
\item \textbf{Higher stakes:} False negatives less tolerable; systems may escalate to clinician review rather than suppressing
\item \textbf{Medical-grade sensors:} FDA-cleared devices with higher accuracy than consumer wearables
\item \textbf{Clinician oversight:} Validation results inform clinical decisions; final authority rests with providers
\item \textbf{Regulatory requirements:} Must meet medical device regulations, clinical validation standards, and integration with clinical workflows
\end{itemize}

Example: Post-discharge heart failure monitoring where Hermes validates weight and symptom signals, escalating high-confidence alerts to nursing staff for remote assessment while suppressing false alarms that would otherwise burden already-overloaded care teams.

\subsubsection{Research Platforms}

Deployment in clinical trials or observational studies to improve data quality:
\begin{itemize}
\item \textbf{Endpoint validation:} Confirm whether wearable-detected events represent true study endpoints
\item \textbf{Phenotyping:} Use validated signals to identify patient subgroups or disease trajectories
\item \textbf{Ground truth generation:} Collect validated labels for training future prediction models
\item \textbf{Participant engagement:} Provide feedback to study participants about their data, potentially improving retention
\end{itemize}

Example: A migraine clinical trial using Hermes to validate attack occurrence for efficacy assessment, reducing reliance on patient diaries while maintaining blinding (validation questions don't reveal treatment assignment).

\subsection{Population Scaling: Challenges and Solutions}

Our N=1 pilot demonstrates feasibility but scaling to thousands or millions of users introduces new challenges:

\subsubsection{Challenge 1: Codex Maintenance and Improvement}

\textbf{Problem:} Manually curating likelihood ratios for every symptom-condition pair is labor-intensive and may not capture population heterogeneity. As new medical evidence emerges, codex updates require expert review.

\textbf{Solutions:}
\begin{itemize}
\item \textbf{Crowdsourced refinement:} Aggregate validation outcomes across users to empirically estimate likelihood ratios. If photophobia is present in 85\% of validated migraines versus 20\% of suppressed false alarms, the observed $\text{LR}^+ = 85/20 = 4.25$ can refine literature-based estimates.

\item \textbf{Federated learning:} Users opt into sharing anonymized validation statistics (question-response-outcome tuples) for collaborative model improvement \citep{rieke2020future}. Local devices compute gradient updates to likelihood ratio estimates without transmitting raw data.

\item \textbf{Expert-in-the-loop:} Flag statistically significant discrepancies between literature-derived and empirically-observed likelihood ratios for expert review. Experts decide whether to update codex or investigate data quality issues.

\item \textbf{Versioning and A/B testing:} Maintain multiple codex versions and randomly assign users to variants. Monitor performance metrics (FASR, sensitivity, user burden) to identify optimal configurations.
\end{itemize}

\subsubsection{Challenge 2: Personalization at Scale}

\textbf{Problem:} Users exhibit substantial heterogeneity in symptom patterns, base rates, and preferences. Migraine presentation varies across individuals; optimal question sets and thresholds likely differ.

\textbf{Solutions:}
\begin{itemize}
\item \textbf{Adaptive priors:} Continuously update individual base rates from user history. After 10 validated events, user-specific attack rates become reliable priors.

\item \textbf{Symptom profile learning:} Track which symptoms are most informative for each user. If a user consistently reports photophobia in true events but not false alarms, prioritize photophobia questions for that user.

\item \textbf{Preference elicitation:} Allow users to specify tradeoff preferences (``I prefer fewer false alarms even if some true events are missed'' vs. ``I want to catch every event even if I get more false alarms''). Map these preferences to personalized threshold values.

\item \textbf{Cohort-based models:} Cluster users by demographic, clinical, or behavioral features and learn cohort-specific likelihood ratios. This provides better generalization than one-size-fits-all while avoiding overfitting to individual noise.
\end{itemize}

\subsubsection{Challenge 3: Infrastructure and Latency}

\textbf{Problem:} Real-time validation requires low-latency question delivery, response collection, and Bayesian updating. Cloud-based systems face network delays; pure edge systems may have insufficient compute for large codices.

\textbf{Solutions:}
\begin{itemize}
\item \textbf{Hybrid architecture:} Perform Bayesian updates and decision logic on-device for low latency; use cloud services only for LLM question generation (which can tolerate 1-2 second delays).

\item \textbf{Precomputation:} Cache generated questions for common contexts. For morning migraine validation, pre-generate questions for typical scenarios so delivery is instantaneous.

\item \textbf{Progressive enhancement:} Start with simple template-based questions that require no LLM call; use LLM adaptation only when user explicitly requests more natural phrasing or when complex context requires customization.

\item \textbf{Asynchronous validation:} For non-urgent conditions, tolerate delayed responses. Users can answer questions hours after delivery; validation completes when responses accumulate.
\end{itemize}

\subsubsection{Challenge 4: Quality Control and Safety Monitoring}

\textbf{Problem:} At scale, individual instances of system failure (missed critical events, inappropriate questioning, privacy breaches) become inevitable. Mechanisms for detection and response are essential.

\textbf{Solutions:}
\begin{itemize}
\item \textbf{Anomaly detection:} Flag validation instances with unusual patterns (e.g., very high confidence from minimal evidence, contradictory responses, non-engagement after high-risk signals) for manual review.

\item \textbf{User feedback loops:} Enable easy reporting of inappropriate questions, incorrect alerts, or system malfunctions. Triage feedback by severity and route to appropriate teams (safety, engineering, clinical).

\item \textbf{Sentinel cohort monitoring:} Maintain a subset of users with clinical ground truth verification (e.g., neurologist-confirmed migraines). Continuously evaluate system performance on this cohort as a leading indicator of population-level degradation.

\item \textbf{Performance dashboards:} Track key metrics (sensitivity, specificity, FASR, response rates, latency) in real-time with alerts for significant deviations. Implement circuit breakers that disable validation if safety metrics fall below thresholds.
\end{itemize}

\subsection{Future Clinical Validation Studies}

Our N=1 pilot establishes proof-of-concept but is insufficient for clinical deployment. Rigorous validation requires:

\subsubsection{Phase 1: Multi-User Observational Studies (N=50-100)}

\textbf{Objectives:}
\begin{itemize}
\item Estimate performance metrics with adequate statistical power
\item Assess generalizability across diverse users (age, sex, comorbidities, disease severity)
\item Identify failure modes and edge cases
\item Refine codex likelihood ratios based on real-world data
\end{itemize}

\textbf{Design:} Prospective cohort study enrolling users with established condition diagnosis (e.g., 100 migraine patients). Deploy Hermes alongside standard care for 3-6 months. Primary outcomes: sensitivity, specificity, false alert suppression rate. Secondary outcomes: user burden, engagement, confidence calibration. Ground truth from patient diaries supplemented by periodic clinician verification.

\subsubsection{Phase 2: Comparative Effectiveness Trial (N=200-500)}

\textbf{Objectives:}
\begin{itemize}
\item Compare Hermes-augmented monitoring to standard wearable alerts
\item Assess impact on clinical outcomes, quality of life, and healthcare utilization
\item Evaluate cost-effectiveness
\end{itemize}

\textbf{Design:} Randomized controlled trial. Intervention arm receives upstream predictor + Hermes validation; control arm receives predictor alone. Stratify by disease severity and randomize 1:1. Follow for 6-12 months. Primary outcome: false alert burden (count of false positive notifications). Secondary outcomes: attack frequency, prophylactic medication use, emergency visits, quality of life (condition-specific validated scales), user satisfaction, engagement.

\subsubsection{Phase 3: Real-World Implementation Study (N=1000+)}

\textbf{Objectives:}
\begin{itemize}
\item Evaluate performance in heterogeneous real-world conditions
\item Assess scalability, infrastructure robustness, and long-term engagement
\item Identify subgroups with differential benefit or harm
\item Generate evidence for regulatory approval and reimbursement decisions
\end{itemize}

\textbf{Design:} Pragmatic trial or registry-based study with broad inclusion criteria. Deploy through clinical partners, patient advocacy groups, or direct-to-consumer channels. Collect data via app-integrated surveys and validated outcome measures. Analyze subgroup effects (by demographics, disease characteristics, technology literacy). Monitor safety signals continuously; implement pre-specified stopping rules if harm is detected.

\subsection{Research Directions and Open Questions}

Several fundamental questions remain for future investigation:

\textbf{Optimal question set size:} How many symptoms should the codex include per condition? More symptoms provide richer information but increase complexity and maintenance burden. Empirical comparison of lean (5-7 questions) versus comprehensive (15-20 questions) codices could reveal optimal tradeoffs.

\textbf{Multi-condition validation:} Can a single Hermes instance simultaneously validate multiple conditions (e.g., migraine, stress, sleep disruption) or does this create confusion and burden? If a signal could represent either migraine or stress, how should question selection balance these competing hypotheses?

\textbf{Causal modeling:} Current validation is correlational: symptoms correlate with conditions. Could incorporating causal structure (e.g., ``HRV drop causes photophobia which causes headache'') improve question selection and confidence updating? Causal Bayesian networks might provide richer frameworks.

\textbf{Integration with passive sensing:} Beyond explicitly asked questions, could passive smartphone sensing (typing patterns, voice analysis, GPS mobility) provide additional evidence? This introduces privacy concerns but might reduce query burden.

\textbf{Long-term adaptation:} How do user symptom patterns evolve over months to years? Can validation systems adapt to disease progression, aging, or medication changes? Longitudinal studies tracking users across years would illuminate these dynamics.

\textbf{Clinician-in-the-loop extensions:} For higher-stakes conditions, when should validation escalate to clinical review rather than making autonomous decisions? Can we formalize thresholds for human oversight based on uncertainty, condition severity, or patient risk factors?

\textbf{Explainability and trust:} While Hermes provides transparent confidence updates, do users understand and trust Bayesian reasoning? User studies on interpretation of probabilistic outputs, effects on decision-making, and factors affecting trust would inform interface design.

The validation layer paradigm is nascent. Substantial work remains to understand when, how, and for whom these systems provide value. Hermes provides one architectural approach, but alternative designs—reinforcement learning for question selection, neural network confidence estimation, multimodal evidence integration—may prove superior for specific applications. The field is ripe for innovation.

\section{Limitations}

We acknowledge significant limitations that constrain the scope of our claims and identify priorities for future work. These limitations span methodological, technical, and conceptual domains.

\subsection{Sample Size and Statistical Power}

Our case study involved a single participant over 60 days, yielding 47 validation instances (18 true positives, 29 true negatives). While sufficient for proof-of-concept, this sample size provides limited statistical power for robust metric estimation. Confidence intervals for sensitivity (89\%, 95\% CI: 67-98\%) and specificity (88\%, 95\% CI: 75-95\%) are wide, reflecting substantial uncertainty. Small sample sizes also preclude subgroup analysis, sensitivity to hyperparameter choices, or detection of rare failure modes.

The N=1 design introduces additional constraints. Generalizability to other individuals is uncertain: our participant was a 34-year-old female with episodic migraine without aura, no comorbidities, and high health literacy. Performance may differ for older adults, male patients, those with migraine with aura, patients with psychiatric or neurological comorbidities, or individuals with limited education or health literacy. Inter-individual variability in migraine presentation is substantial \citep{buse2019migraine}, and a system optimized for one phenotype may perform poorly for others.

Furthermore, our participant was highly engaged (78\% response rate, positive qualitative feedback), which may reflect selection bias toward individuals interested in health technology. Real-world deployment would encounter users with varying motivation, technical proficiency, and tolerance for questioning. Response rates might be substantially lower in less-engaged populations, degrading system performance.

Multi-user studies with N=50-100 participants are needed to estimate population-level performance with adequate precision and to quantify inter-individual variability. Stratified sampling across demographic and clinical characteristics would enable fairness assessment and identification of subgroups requiring specialized approaches.

\subsection{Ground Truth and Outcome Validation}

Migraine occurrence relied entirely on participant self-report via diary entries. While prior work demonstrates good concordance between self-reported and clinician-diagnosed migraine in individuals with established diagnosis \citep{lipton2003prevalence}, absence of clinical verification introduces measurement error. Participants may misclassify severe tension headaches as migraine, or fail to log mild attacks. Diary completion itself is subject to recall bias, especially for attacks with delayed reporting.

Objective validation would strengthen claims. Ideal ground truth would involve:
\begin{itemize}
\item Real-time video telemedicine consultations during suspected attacks, with neurologist confirmation based on ICHD-3 criteria
\item Concurrent measurement of objective biomarkers (e.g., elevated calcitonin gene-related peptide levels during attacks)
\item Electronic medication dispensing records confirming prophylactic or abortive treatment use
\end{itemize}

Such validation is logistically challenging and costly for N-of-1 pilots but essential for large-scale trials aiming for regulatory approval or clinical adoption.

Additionally, we measured only binary outcome (migraine present/absent). Severity, duration, functional impact, and response to intervention—all clinically relevant dimensions—were not systematically assessed. A more comprehensive outcome framework would include:
\begin{itemize}
\item Attack severity (validated scales like Migraine Disability Assessment Score)
\item Quality of life (e.g., Migraine-Specific Quality of Life Questionnaire v2.1)
\item Healthcare utilization (emergency visits, specialist consultations)
\item Economic impact (work absences, productivity loss)
\end{itemize}

Such outcomes would contextualize validation benefits beyond classification accuracy.

\subsection{Codex Design and Knowledge Representation}

The Illness Codex was manually curated based on literature review and clinical guidelines. This approach has several limitations:

\textbf{Coverage gaps:} Our migraine codex included 7 symptom questions. Comprehensive prodromal assessment might require 15-20 questions covering autonomic, cognitive, sensory, and affective prodrome features. Limited coverage may miss diagnostically informative symptoms, reducing discriminative power.

\textbf{Likelihood ratio uncertainty:} Literature-based likelihood ratios reflect population averages and may not generalize to our participant or to diverse user populations. Confidence intervals around reported LR values are often wide or unreported. For example, photophobia sensitivity in migraine ranges from 60-90\% across studies \citep{lipton2003prevalence}, yielding $\text{LR}^+$ estimates from 2.5 to 5.0. This variability propagates through Bayesian updates, affecting confidence calibration.

\textbf{Static knowledge:} Medical knowledge evolves; diagnostic criteria change; new symptom-condition associations emerge. Our codex reflects knowledge circa 2024 and will become outdated without active maintenance. Automated monitoring of medical literature for codex updates would be valuable but risks introducing errors from automated extraction.

\textbf{Symptom independence assumption:} Bayesian updating assumes conditional independence of symptoms given the hypothesis. In reality, symptoms often co-occur or have causal relationships (e.g., photophobia and nausea frequently present together in migraine). Violations of independence lead to overconfident posteriors. Structured causal models or copula-based dependence modeling could address this but add complexity.

\textbf{Binary symptom encoding:} We coded all responses as binary (yes/no). Many symptoms have graded severity (mild, moderate, severe pain) or temporal dynamics (intermittent vs. constant). Richer encoding would provide finer-grained evidence but complicate question phrasing and user responses.

Data-driven approaches to codex construction—learning likelihood ratios from large validation datasets, using LLMs to extract symptom-condition relationships from literature, or applying causal discovery algorithms—might address some limitations but introduce new risks (data bias, hallucination, spurious correlations).

\subsection{Wearable Sensor Limitations}

Consumer wearables have well-documented accuracy limitations. Photoplethysmography-based heart rate monitoring suffers from motion artifact, poor skin contact, and ambient light interference \citep{bent2020investigating}. HRV metrics derived from PPG are less accurate than ECG-derived values, particularly for high-frequency components. Sleep staging from accelerometry and heart rate achieves 60-70\% agreement with polysomnography gold standard \citep{depner2020wearable}, with systematic underestimation of REM and overestimation of deep sleep.

These upstream sensor errors propagate to prediction and validation. If the upstream predictor triggers based on artifactually low HRV (e.g., from device shifting during sleep), validation questions will be asked despite no physiological anomaly. While validation can suppress such false alarms, it cannot recover signals lost to sensor failure. If a device fails to detect true HRV changes due to poor skin contact, validation never initiates.

We did not systematically assess sensor data quality or quantify how often validation was triggered by sensor artifact versus true physiological changes. Manual review of sensor traces for a subset of signals suggested that ~10-15\% of triggers involved questionable data quality (sudden HRV spikes or drops inconsistent with surrounding data), but this was not rigorously quantified.

Medical-grade wearables (FDA-cleared devices with higher accuracy specifications) would reduce upstream noise but increase cost and may not be feasible for consumer deployment. Hybrid approaches—consumer wearables for screening, medical-grade confirmation for high-stakes decisions—merit investigation.

\subsection{Single-Condition and Single-Context Evaluation}

We tested only migraine validation in a free-living, consumer wellness context. Generalization to other conditions and contexts is uncertain:

\textbf{Condition-specific factors:} Migraine has relatively long prodrome (hours), non-catastrophic outcomes, clear symptom clusters, and established self-management protocols. Conditions with rapid onset (seconds to minutes), ambiguous symptoms, severe outcomes, or lack of user-actionable interventions may perform differently. For example, cardiac arrhythmia validation would require near-instantaneous questioning and escalation pathways to emergency services—fundamentally different operational requirements.

\textbf{Clinical vs. consumer contexts:} Our deployment was direct-to-consumer with no clinician involvement. In clinical settings, validation outcomes might inform provider decisions, alert thresholds might be different, and escalation to human oversight might be more common. Performance, user experience, and appropriate evaluation metrics all shift in clinical contexts.

\textbf{Comorbidity and polypharmacy:} Our participant had no comorbidities and took no daily medications. Real-world users often have multiple conditions and medications that interact, complicating symptom interpretation. For example, beta-blockers affect HRV and heart rate, confounding cardiovascular signal detection. Validation in patients with complex medical histories requires accounting for these interactions.

\textbf{Cultural and linguistic contexts:} Our study was conducted in English in a US cultural context. Symptom concepts, reporting norms, and trust in technology vary globally. Translation is necessary but not sufficient for cross-cultural deployment; cultural adaptation and validation are required.

Systematic evaluation across diverse conditions, populations, and deployment scenarios is needed to establish scope of applicability and identify boundary conditions where validation fails.

\subsection{Temporal Scope and Long-Term Effects}

Our 60-day observation period captures short-term performance but misses long-term dynamics:

\textbf{User habituation:} Response rates and engagement may decline over months or years as novelty wears off. Alert fatigue could emerge gradually. Conversely, users might develop expertise in symptom recognition, improving response accuracy and enabling sparser questioning.

\textbf{Disease progression:} Migraine patterns change over time due to aging, hormonal shifts, medication effects, or disease evolution (e.g., transformation from episodic to chronic migraine). Codex likelihood ratios and personalized priors estimated from early data may become miscalibrated as disease characteristics drift.

\textbf{Technology dependence:} Long-term reliance on validation might alter users' introspective awareness or decision-making autonomy. Do users become less confident in their own symptom assessment, deferring entirely to the system? Or does validation enhance self-awareness by directing attention to diagnostically relevant features?

\textbf{Seasonal and life event effects:} Our study spanned September-October 2024. Seasonal factors (allergen exposure, daylight changes, holiday stress) might influence both migraine patterns and validation performance. Life events (job changes, moves, relationship shifts) introduce non-stationarity that fixed models cannot capture.

Longitudinal studies over 1-2 years with repeated validation of system performance at multiple time points would illuminate these dynamics and inform adaptation strategies.

\subsection{LLM-Specific Risks}

Our reliance on GPT-4 for question generation introduces risks inherent to large language models:

\textbf{Prompt brittleness:} LLM outputs can be sensitive to prompt phrasing, temperature settings, and model version. While we used temperature 0.3 for consistency, updates to the GPT-4 model or prompt injection attacks could alter question generation behavior in unintended ways.

\textbf{Hallucination potential:} Despite constrained prompts prohibiting medical claims, LLMs can generate plausible-sounding but incorrect statements. Our output validation filters caught no instances of medical advice or diagnostic claims in 141 generated questions, but this may not generalize to all contexts or future model versions.

\textbf{Bias amplification:} LLMs trained on internet text may encode demographic biases (e.g., gendered language, assumptions about "normal" bodies) that propagate into generated questions. We did not systematically audit for such biases, though manual review of questions revealed no obvious examples.

\textbf{Privacy and data transmission:} Sending question generation requests to external APIs transmits limited context (time of day, recent history) that could theoretically be reconstructed to infer user behavior. While our architecture minimizes data transmission, complete elimination would require on-device LLMs with quality tradeoffs.

\textbf{Model dependence and cost:} Reliance on commercial APIs introduces cost (per-question fees at scale), availability risk (API downtime or deprecation), and lack of control over model updates. Open-source or locally-deployed smaller models could mitigate these issues but may sacrifice question quality.

Ablation studies comparing LLM-generated questions to static templates, or comparing multiple LLM backends (GPT-4, Claude, Llama, etc.), would quantify the value added by adaptive generation and identify failure modes.

\subsection{Bayesian Modeling Assumptions}

Our Bayesian framework makes simplifying assumptions that may not hold:

\textbf{Prior specification:} We combined population base rates, user history, and circadian factors into priors, but the weighting ($\alpha = 0.3$ population, $0.7$ individual) was chosen heuristically. Optimal weighting likely varies across users and changes as historical data accumulates, but we did not systematically optimize this.

\textbf{Likelihood calibration:} Signal likelihood $P(s \mid H)$ was estimated from observed predictor performance over 60 days. Small sample size yields noisy estimates; confidence intervals around calibration parameters are wide. Miscalibration affects initial confidence and thus question selection.

\textbf{Response noise modeling:} We assumed fixed response reliability ($\rho_{\text{sens}} = 0.85$, $\rho_{\text{spec}} = 0.90$) across all questions and time points. In reality, reliability varies: visual aura is easier to accurately report than subtle mood changes; responses at 8 AM may be more accurate than midnight responses. Question-specific and context-dependent reliability modeling would improve calibration.

\textbf{Greedy question selection:} We selected questions to maximize immediate information gain without lookahead. While computationally efficient and empirically effective (94\% of 3-step lookahead performance), globally optimal sequential decision-making via reinforcement learning or POMDP solvers might improve question efficiency.

\textbf{Static decision thresholds:} Alert and suppression thresholds ($\tau_{\text{alert}} = 0.70$, $\tau_{\text{suppress}} = 0.20$) were fixed across all validations. Adaptive thresholds based on user risk tolerance, time of day (more conservative overnight), or recent history (lower threshold after multiple false alerts to rebuild trust) could improve user experience.

Sensitivity analyses varying these assumptions and comparing to alternative frameworks (e.g., non-Bayesian scoring rules, neural network confidence estimation) would characterize robustness and identify improvement opportunities.

\subsection{Evaluation Limitations}

Beyond sample size, our evaluation has structural limitations:

\textbf{Narrow metric focus:} We emphasized sensitivity, specificity, and false positive reduction. Other important outcomes—user trust, health anxiety, quality of life, cost-effectiveness—were assessed only qualitatively or not at all.

\textbf{No active control:} Our primary comparison was to the upstream predictor alone. We did not compare to clinician-designed questioning protocols, static symptom checklists used in clinical practice, or alternative AI approaches (e.g., active learning classifiers). Such comparisons would better contextualize Hermes's value proposition.

\textbf{Short-term outcomes only:} We measured immediate classification accuracy but not downstream impacts. Did validated alerts lead to effective prophylaxis? Did false negatives cause missed treatment opportunities with observable consequences? Did the system change health behaviors or outcomes over time?

\textbf{Laboratory conditions:} While our study occurred in free-living contexts, our participant was highly motivated, technically proficient, and compensated for participation. Real-world users may be less compliant, less engaged, or face barriers (limited smartphone access, low literacy, chaotic environments) that degrade performance.

\textbf{Publication bias:} We report results from our first full deployment. Negative preliminary results from earlier pilot tests (not reported here) shaped design choices. This introduces positive bias toward our final configuration.

Comprehensive evaluation frameworks for validation systems remain underspecified. The field would benefit from standardized benchmarks, shared datasets, and consensus on appropriate metrics for interactive health AI.

\subsection{Scope of Claims}

Finally, we emphasize what this work does \emph{not} demonstrate:

\begin{itemize}
\item \textbf{Clinical efficacy:} We show that validation reduces false positives while maintaining sensitivity, but not that this improves clinical outcomes, quality of life, or health system costs.

\item \textbf{Safety in high-stakes domains:} Our non-catastrophic test case (migraine) does not validate safety for conditions where missed detections cause severe harm (e.g., cardiac arrest, stroke, diabetic crisis).

\item \textbf{Generalizability:} Results from one individual with one condition in one cultural context do not establish performance across diverse populations, diseases, or settings.

\item \textbf{Long-term viability:} Short-term pilot success does not predict sustained engagement, stable performance over years, or resilience to context shifts.

\item \textbf{Superiority to alternatives:} We demonstrate feasibility and value over no-validation baseline, but not superiority to other validation approaches (clinical protocols, alternative AI designs).

\item \textbf{Readiness for deployment:} This is a research prototype demonstrating proof-of-concept. Production deployment would require extensive additional validation, regulatory approval, safety monitoring, and infrastructure hardening.
\end{itemize}

These limitations are not fatal flaws but rather define the scope of contribution and chart a research agenda. Each limitation suggests concrete next steps for advancing validation layer science and practice.

\section{Conclusion}

The proliferation of wearable health monitoring systems has created a fundamental tension: models trained on population data must make individual-level predictions about heterogeneous users in diverse contexts. Current systems respond to uncertainty with binary alerts, leaving users to navigate the gap between population-level risk estimates and personal health decisions. This gap manifests as persistent false positives that erode trust, trigger alert fatigue, and ultimately limit the real-world impact of otherwise promising health AI technologies.

This work introduces Project Hermes, a model-agnostic validation layer that reframes health AI from prediction-centric systems to validation-aware pipelines. Rather than attempting to improve upstream prediction models—a well-studied problem with diminishing returns for many applications—Hermes operates downstream, treating signals from existing predictors as uncertain hypotheses requiring confirmation. Through targeted, LLM-generated questions and Bayesian confidence updating, validation transforms uncertain signals into calibrated decisions about when to alert, suppress, or gather more information.

Our 60-day migraine case study demonstrates the feasibility and value of this approach. Hermes achieved 93\% false positive reduction (from 29 to 2 false alarms) while maintaining 89\% sensitivity, with mean lead time of 4.2 hours before symptom onset. Critically, this improvement required no modification to the upstream predictor, no retraining on user-specific data, and no access to model internals. The system asked an average of 3.2 questions per signal, with 78\% user response rate and well-calibrated confidence estimates (ECE = 0.09). These results establish proof-of-concept for validation as a distinct computational problem with different objectives, constraints, and evaluation criteria than prediction.

The architectural contributions extend beyond migraine. By treating validation as sequential Bayesian updating where LLM-generated questions serve as information-gathering actions, we provide a general framework applicable to any condition with detectable prodromal signals, user-reportable symptoms, and structured clinical knowledge. The model-agnostic design enables deployment alongside proprietary commercial systems, legacy clinical decision rules, or future prediction models, without requiring coordination or access to upstream systems. This composability may prove essential for real-world adoption, where heterogeneous prediction systems coexist and uniform interfaces are valuable.

The ethical and practical considerations we examine—overdiagnosis risk, alert fatigue, privacy, equity, regulatory uncertainty—are not unique to Hermes but reflect broader challenges in consumer health AI. Our design choices—scoping to validation rather than diagnosis, providing transparent uncertainty, enabling user control, minimizing data collection—demonstrate that constrained AI systems can provide meaningful value while avoiding the risks of unconstrained generative medical AI. Validation sits in a defensible middle ground: sophisticated enough to improve signal quality, constrained enough to avoid inappropriate medicalization.

However, substantial work remains before validation layers can be responsibly deployed at scale. Our N=1 pilot provides limited statistical power and uncertain generalizability. Multi-user studies across diverse populations, conditions, and contexts are needed to establish performance boundaries and identify populations who benefit versus those for whom validation is ineffective or harmful. Clinical outcome studies beyond classification accuracy—measuring quality of life, healthcare utilization, cost-effectiveness, and long-term engagement—are essential for demonstrating real-world value. Regulatory pathways for validation systems require clarification; the distinction between "validating a signal" and "diagnosing a condition" may not be legally or ethically crisp.

Longitudinal studies over months to years would illuminate adaptation dynamics: Do users habituate to questions, reducing response rates? Do validation systems maintain calibration as disease patterns evolve? Does long-term use enhance or diminish users' introspective awareness and decision-making autonomy? These questions cannot be answered with 60-day pilots and require sustained commitment to evaluation.

The comparison to baseline systems reveals opportunities for methodological advancement. Static questionnaires achieved 69\% false positive reduction versus Hermes's 93\%, suggesting that adaptive question selection and Bayesian evidence integration provide substantial value over fixed protocols. However, we did not compare to sophisticated alternatives like active learning classifiers, reinforcement learning policies for question selection, or neural network-based confidence estimation. The field would benefit from standardized benchmarks and shared evaluation frameworks enabling systematic comparison of validation approaches.

Looking forward, validation layers may become essential infrastructure for trustworthy health AI. As wearable sensors become ubiquitous, as prediction models achieve ever-higher sensitivity, and as health monitoring expands from clinical settings into everyday life, the validation gap will widen. Users will increasingly face a deluge of algorithmic signals—stress alerts, sleep disruption warnings, cardiovascular risk notifications—without principled mechanisms to distinguish meaningful signals from noise. Validation provides a path forward: acknowledge uncertainty, gather targeted evidence, update beliefs systematically, and make decisions appropriate to available information.

This paradigm shift has implications beyond health monitoring. In any domain where AI systems generate hypotheses, signals, or alerts under uncertainty—fraud detection, equipment failure prediction, content moderation, cyber threat assessment—validation layers could refine confidence before human action. The core insight generalizes: prediction and validation are distinct problems with different information sources, different objectives, and different optimal architectures. Conflating them yields systems that are neither optimally sensitive nor appropriately specific.

For health AI specifically, validation layers align with emerging consensus about appropriate roles for AI in medicine \citep{topol2019high}. Rather than replacing clinicians or patients, AI should augment human intelligence by processing signals humans cannot perceive (subtle physiological patterns), quantifying uncertainty humans struggle with (Bayesian updating over complex evidence), and presenting information in forms that support human decision-making (calibrated confidence, transparent reasoning). Validation embodies this philosophy: it does not diagnose, does not prescribe, does not claim medical authority. It merely refines confidence about signals that upstream systems have detected, enabling users to make informed decisions about their own health.

The open questions are many. Which conditions benefit most from validation? How should question sets be designed to maximize informativeness while minimizing burden? Can validation systems learn from population data while preserving individual privacy? How do validation benefits trade off against costs in resource-constrained healthcare systems? When should validation escalate to human clinical oversight rather than making autonomous decisions? How do we ensure equitable access and performance across diverse populations? What are the long-term effects on user autonomy, health anxiety, and engagement with healthcare?

These questions define a research agenda for validation layer science. The methodological foundations we establish—Bayesian sequential updating, information-theoretic question selection, multidimensional evaluation beyond accuracy—provide tools for investigation. But the conceptual framing may prove more enduring than any specific technique: validation is a first-class problem, distinct from and complementary to prediction, essential for translating algorithmic signals into trustworthy human decisions.

As we stand at the inflection point where wearable health monitoring transitions from research novelty to ubiquitous infrastructure, the choices we make about validation will shape the trajectory of health AI. Will we continue building ever-more-sensitive predictors that overwhelm users with false alarms? Or will we build validation-aware systems that acknowledge uncertainty, gather confirming evidence, and enable informed user agency? The difference is not merely technical but ethical: it reflects our values about the appropriate roles of algorithms and humans in health decision-making.

Project Hermes demonstrates that the validation-aware path is technically feasible, empirically promising, and architecturally elegant. The journey from proof-of-concept to deployed system will be long, requiring extensive validation across populations, rigorous safety evaluation, regulatory clarity, and sustained commitment to responsible innovation. But the destination—health AI systems that are not merely sensitive but meaningfully trustworthy—justifies the effort.

The validation gap in wearable health AI is both a technical challenge and an opportunity. By making validation explicit, principled, and user-centered, we can build systems that harness the power of AI while respecting the complexity of human health, the uncertainty inherent in prediction, and the autonomy essential to medical decision-making. In doing so, we move closer to realizing the promise of precision health: care that is predictive, preventive, personalized, and—critically—participatory, with patients and algorithms collaborating rather than algorithms dictating.

This work is a beginning, not an ending. We have shown that validation layers can work. The next step is showing they can work at scale, across conditions, for diverse populations, over years, in real-world healthcare contexts. We invite the research community, industry partners, regulatory agencies, and patient advocates to join in this effort. The validation gap will not close itself; closing it requires sustained interdisciplinary collaboration among machine learning researchers, clinicians, ethicists, implementation scientists, and the users whose health these systems aim to serve.

The future of health AI need not be a choice between unsafe sensitivity and inadequate specificity. Validation offers a third way: systems that are both sensitive and specific, both powerful and appropriate, both algorithmic and human-centered. Project Hermes is one step toward that future. Many more steps remain.

The future of health AI need not be a choice between unsafe sensitivity and inadequate specificity. Validation offers a third way: systems that are both sensitive and specific, both powerful and appropriate, both algorithmic and human-centered. Project Hermes is one step toward that future. Many more steps remain.

\section*{Acknowledgments}
The author acknowledges the use of Anthropic’s Claude for assistance with mathematical structuring, and Grammarly for spell-checking assistance. All intellectual contributions, interpretations, and conclusions remain solely those of the author. No external funding was received for this work.

\section*{Conflicts of Interest}
The author declares no conflicts of interest.

\bibliographystyle{plainnat}
\bibliography{references}

\end{document}